\documentclass[twoside,11pt]{article}

\usepackage{jmlr2e}

\RequirePackage{hyperref}
\RequirePackage{amsmath}
\RequirePackage{subfigure}
\RequirePackage{commath}
\RequirePackage{multirow}
\RequirePackage{lscape}
\RequirePackage{cleveref}
\newcommand{\indicator}{\mathbf{1}}
\newcommand{\GP}{\text{Gauss}}
\newcommand{\NGP}{\text{NegBin}}
\newcommand{\iid}[1]{\stackrel{\text{iid}}{#1}}

\jmlrheading{VOLUME}{YEAR}{PAGES}{DATE}{DATE}{Matthew J. Johnson and Alan S. Willsky}

\ShortHeadings{Bayesian Nonparametric Hidden Semi-Markov Models}{Johnson and Willsky}
\firstpageno{1}

\begin{document}

\title{Bayesian Nonparametric Hidden Semi-Markov Models }

\author{\name Matthew J. Johnson \email mattjj@csail.mit.edu \\
    \addr
    Laboratory for Information and Decision Systems\\
    Department of EECS\\
    Massachusetts Institute of Technology\\
    Cambridge, MA 02139-4307, USA
    \AND
    \name Alan S. Willsky \email willsky@mit.edu \\
    \addr
    Laboratory for Information and Decision Systems\\
    Department of EECS\\
    Massachusetts Institute of Technology\\
    Cambridge, MA 02139-4307, USA
    }

\maketitle

\begin{abstract}
    There is much interest in the Hierarchical Dirichlet Process Hidden Markov
Model (HDP-HMM) as a natural Bayesian nonparametric extension of the ubiquitous
Hidden Markov Model for learning from sequential and time-series data. However,
in many settings the HDP-HMM's strict Markovian constraints are undesirable,
particularly if we wish to learn or encode non-geometric state durations. We
can extend the HDP-HMM to capture such structure by drawing upon
explicit-duration semi-Markov modeling, which has been developed mainly in the
parametric non-Bayesian setting, to allow construction of highly interpretable
models that admit natural prior information on state durations.

In this paper we introduce the explicit-duration Hierarchical Dirichlet Process
Hidden semi-Markov Model (HDP-HSMM) and develop sampling algorithms for
efficient posterior inference. The methods we introduce also provide new
methods for sampling inference in the finite Bayesian HSMM\@.
Our modular Gibbs sampling methods can be embedded in samplers for larger
hierarchical Bayesian models, adding semi-Markov chain modeling as another tool
in the Bayesian inference toolbox.  We demonstrate the utility of the HDP-HSMM
and our inference methods on both synthetic and real experiments.

\end{abstract}


\section{Introduction}
\label{sec:intro}
Given a set of sequential data in an unsupervised setting, we often aim to
infer meaningful states, or ``topics,'' present in the data along with
characteristics that describe and distinguish those states. For example, in a
speaker diarization (or who-spoke-when) problem, we are given a single audio
recording of a meeting and wish to infer the number of speakers present, when
they speak, and some characteristics governing their speech patterns
\citep{tranter2006overview,emilysticky}. Or in separating a home power
signal into the power signals of individual devices, we would be able to
perform the task much better if we were able to exploit our prior knowledge
about the levels and durations of each device's power modes
\citep{kolter2011redd}.  Such learning problems for sequential data are
pervasive, and so we would like to build general models that are both flexible
enough to be applicable to many domains and expressive enough to encode the
appropriate information.

Hidden Markov Models (HMMs) have proven to be excellent general models for
approaching learning problems in sequential data, but they have two significant
disadvantages: (1) state duration distributions are necessarily restricted to a
geometric form that is not appropriate for many real-world data, and (2) the
number of hidden states must be set a priori so that model complexity is not
inferred from data in a Bayesian way.

Recent work in Bayesian nonparametrics has addressed the latter issue. In
particular, the Hierarchical Dirichlet Process HMM (HDP-HMM) has provided a
powerful framework for inferring arbitrarily large state complexity from data
\citep{originalhdp,beal2002infinite}. However, the HDP-HMM does not address the
issue of non-Markovianity in real data. The Markovian disadvantage is even
compounded in the nonparametric setting, since non-Markovian behavior in data
can lead to the creation of unnecessary extra states and unrealistically rapid
switching dynamics \citep{emilysticky}.

One approach to avoiding the rapid-switching problem is the Sticky HDP-HMM
\citep{emilysticky}, which introduces a learned global self-transition bias to
discourage rapid switching. Indeed, the Sticky model has demonstrated
significant performance improvements over the HDP-HMM for several applications.
However, it shares the HDP-HMM's restriction to geometric state durations, thus
limiting the model's expressiveness regarding duration structure. Moreover, its
global self-transition bias is shared among all states, and so it does not
allow for learning state-specific duration information. The infinite
Hierarchical HMM \citep{ihhmm} induces non-Markovian state durations at the
coarser levels of its state hierarchy, but even the coarser levels are
constrained to have a sum-of-geometrics form, and hence it can be difficult to
incorporate prior information. Furthermore, constructing posterior samples from
any of these models can be computationally expensive, and finding efficient
algorithms to exploit problem structure is an important area of research.

These potential limitations and needed improvements to the HDP-HMM motivate
this investigation into explicit-duration semi-Markov modeling, which has a
history of success in the parametric (and usually non-Bayesian) setting. We
combine semi-Markovian ideas with the HDP-HMM to construct a general class of
models that allow for both Bayesian nonparametric inference of state complexity
as well as general duration distributions. In addition, the
sampling techniques we develop for the Hierarchical Dirichlet Process Hidden
semi-Markov Model (HDP-HSMM) provide new approaches to inference in HDP-HMMs
that can avoid some of the difficulties which result in slow mixing rates. We
demonstrate the applicability of our models and algorithms on both synthetic
and real datasets.

The remainder of this paper is organized as follows. In
Section~\ref{sec:background}, we describe explicit-duration HSMMs and existing
HSMM message-passing algorithms, which we use to build efficient Bayesian
inference algorithms.  We also provide a brief treatment of the
Bayesian nonparametric HDP-HMM and sampling inference algorithms.
In Section~\ref{sec:models} we develop the HDP-HSMM and related models. In
Section~\ref{sec:algorithms} we develop extensions of the weak-limit and direct
assignment samplers~\citep{originalhdp} for the HDP-HMM to our models and
describe some techniques for improving the computational efficiency in some
settings.

Section~\ref{sec:experiments} demonstrates the effectiveness of the HDP-HSMM on
both synthetic and real data. In synthetic experiments, we demonstrate that our
sampler mixes very quickly on data generated by both HMMs and HSMMs and
accurately learns parameter values and state cardinality. We also show that
while an HDP-HMM is unable to  capture the statistics of an HSMM-generated
sequence, we can build HDP-HSMMs that efficiently learn whether data were
generated by an HMM or HSMM\@. As a real-data experiment, we apply the HDP-HSMM
to a problem in power signal disaggregation.

\section{Background and Notation}
\label{sec:background}
In this section, we outline three main background topics: our notation for
Bayesian HMMs, conventions for explicit-duration HSMMs, and the Bayesian nonparametric HDP-HMM.

\subsection{HMMs}
\subsubsection{Basic Notation}
\begin{figure}
    \centering
    \includegraphics[height = 1.25in]{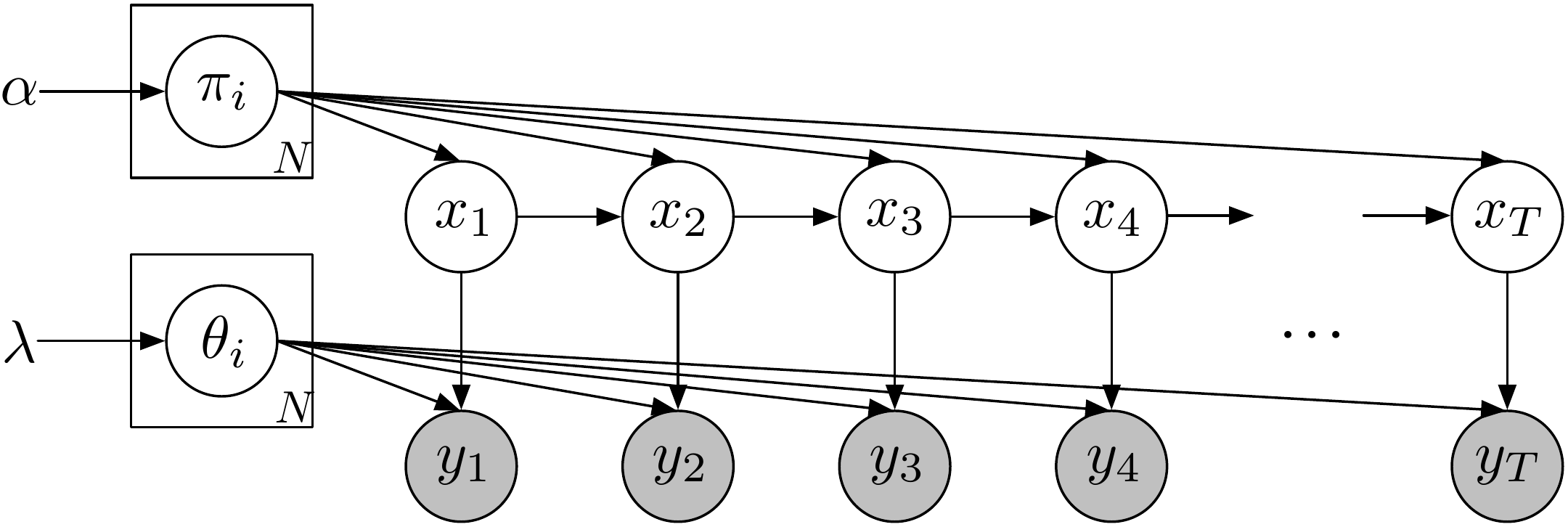}
    \caption{Basic graphical model for the Bayesian HMM\@. Parameters for the
        transition, emission, and initial state distributions are random
        variables. The symbol $\alpha$ represents the hyperparameter for the
        prior distributions on state-transition parameters. The shaded nodes
        indicate observations on which we condition to form the posterior
        distribution over the unshaded latent components.}
    \label{fig:hmm-bayesian}
\end{figure}

The core of the HMM consists of two layers: a layer of hidden \emph{state}
variables and a layer of \emph{observation} or \emph{emission} variables, as
shown in Figure~\ref{fig:hmm-bayesian}.
The hidden state sequence, $x = (x_t)_{t=1}^T$, is a
sequence of random variables on a finite alphabet, i.e. $x_t \in
\{1,2,\ldots,N\}$, that form a Markov chain. In this paper, we focus
on time-homogeneous models, in which the transition distribution does
not depend on $t$. The transition parameters are collected into a
row-stochastic transition matrix $\pi = (\pi_{ij})_{i,j =1}^N$ where $\pi_{ij}
= p(x_{t+1} = j | x_{t} = i)$. We also use $\{\pi_i\}$ to refer to the set of
rows of the transition matrix. We use $p(y_t | x_t, \{\theta_i\})$ to denote
the emission distribution, where $\{\theta_i\}$ represents parameters.

The Bayesian approach allows us to model uncertainty over the parameters and
perform model averaging (e.g.~forming a prediction of an observation $y_{T+1}$
by integrating out all possible parameters and state sequences), generally at
the expense of somewhat more expensive algorithms. This paper is concerned with
the Bayesian approach and so the model parameters are treated as random
variables, with their priors denoted $p(\pi | \alpha)$ and $p(\{\theta_i\} |
H)$.

% \subsubsection{Posterior Inference via Gibbs Sampling} \label{sec:hmm_gibbs} We
% can perform posterior inference in the HMM with a Gibbs sampling algorithm.  An
% iteration of a standard Gibbs sampler samples the following conditional random
% variables: \begin{itemize} \item[] $(x_t)        | \{\theta_i\}, \{\pi_i\},
%         (y_t)$ \item[] $\{\pi_i\}    | \alpha, (x_t)$ \item[] $\{\theta_i\} |
%         H, (x_t), (y_t)$ \end{itemize} Sampling $\{\pi_i\}$ and $\{\theta_i\}$
% from their conditional distributions is a standard problem in Bayesian
% methodology which depends on the specific model distributions chosen; such
% issues are thoroughly described in, e.g., \citep{bishop}. The entire
% conditional state sequence $(x_t)$ can be be block-sampled by passing messages
% backwards, using the familiar HMM message passing scheme to sum over possible
% state sequences, and sampling forwards using the backwards messages; see, e.g.,
% Appendix B.2 of \citep{emilythesis}.

\subsection{HSMMs}
\label{sec:classical_hsmm}
There are several approaches to hidden semi-Markov models
\citep{murphy,yu2010hidden}. We focus on \emph{explicit duration}
semi-Markov modeling; i.e., we are interested in the setting where each state's
duration is given an explicit distribution. Such
HSMMs are generally treated from a non-Bayesian perspective in the literature,
where parameters are estimated and fixed via an approximate maximum-likelihood
procedure (particularly the natural Expectation-Maximization algorithm, which
constitutes a local search).

The basic idea underlying this HSMM formalism is to augment the generative
process of a standard HMM with a random state duration time, drawn from some
state-specific distribution when the state is entered. The state remains
constant until the duration expires, at which point there is a Markov
transition to a new state. We use the random variable $D_t$ to denote the
duration of a state that is entered at time $t$, and we write the probability
mass function for the random variable as $p(d_t | x_t=i)$.

\begin{figure}[tbp] \centering
    \includegraphics[height=1.25in]{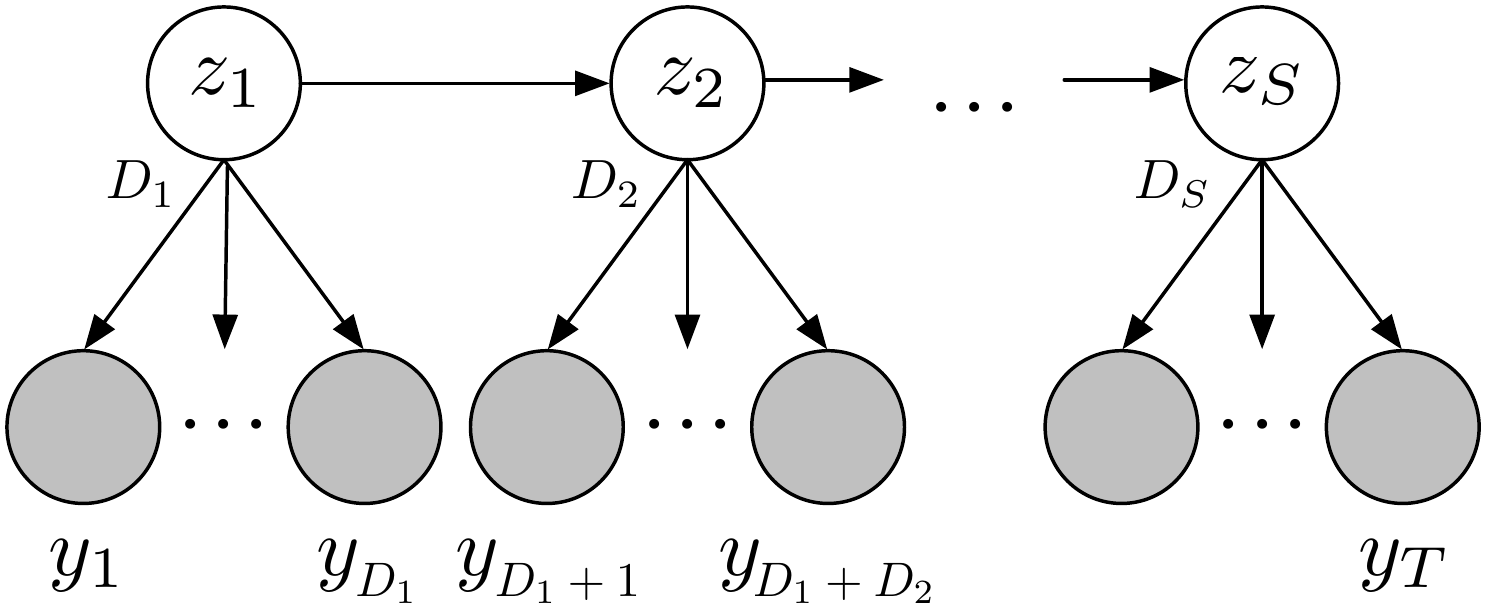}
    \caption{HSMM interpreted as a Markov chain on a set of super-states,
        $(z_s)_{s=1}^S$. The number of shaded nodes associated with each $z_s$,
        denoted by $D_s$, is drawn from a state-specific duration
        distribution.} \label{fig:hsmm_segments} \end{figure}

A graphical model for the explicit-duration HSMM is shown in
Figure~\ref{fig:hsmm_segments} (from \citep{murphy}), though
the number of nodes in the graphical model is itself random. In this picture,
we see there is a Markov chain (without self-transitions) on $S$
``super-state'' nodes, $(z_s)_{s=1}^S$, and these super-states in turn emit
random-length segments of observations, of which we observe the first $T$.
Here, the symbol $D_s$ is used to denote the random length of the observation
segment of super-state $s$ for $s=1,\ldots,S$. The ``super-state'' picture
separates the Markovian transitions from the segment durations.
% , and is helpful
% in building sampling techniques for the generalized models introduced in this
% paper.

When defining an HSMM model, one must also choose whether the observation
sequence ends exactly on a segment boundary or whether the observations are
\emph{censored} at the end, so that the final segment may possibly be cut off
in the observations. We focus on the right-censored formulation in this paper,
but our models and algorithms can easily be modified to the uncensored or
left-censored cases. For a further discussion, see \citep{guedon}.

It is possible to perform efficient message-passing inference along an HSMM
state chain (conditioned on parameters and observations) in a way similar to
the standard alpha-beta dynamic programming algorithm for standard HMMs. The
``backwards'' messages are crucial in the development of efficient sampling
inference in Section~\ref{sec:algorithms} because the message values can be
used to efficiently compute the posterior information necessary to block-sample
the hidden state sequence $(x_t)$, and so we briefly describe the relevant part
of the existing HSMM message-passing algorithm. As derived in \citep{murphy},
we can define and compute the backwards messages\footnote{In \citep{murphy} and
    others, the symbols $\beta$ and $\beta^*$ are used for the messages, but to
    avoid confusion with our HDP parameter $\beta$, we use the symbols $B$ and
    $B^*$ for messages.} $B$ and $B^*$ as follows:

\begin{align}
    \label{eqn:backwards-messages}
    B_t(i) & := p(y_{t+1:T} | x_t = i,
    F_t = 1) \\ &= \sum_{j} B_t^*(j) p(x_{t+1} = j| x_t = i)\\ B_t^*(i) &:=
    p(y_{t+1:T} | x_{t+1} = i, F_t = 1)\\ &= \sum_{d=1}^{T-t} B_{t + d}(i)
    \underbrace{p(D_{t+1}=d|x_{t+1} = i)}_{\text{duration prior term}} \cdot
    \underbrace{p(y_{t+1:t+d} | x_{t+1}=i, D_{t+1}=d)}_{\text{likelihood term}}
    \\ &\quad + \underbrace{p(D_{t+1} > T-t | x_{t+1}=i) p(y_{t+1:T} |
        x_{t+1}=i, D_{t+1}>T-t)}_{\text{censoring term}} \\
    \label{eqn:backwards-messages-end} B_T(i) &:= 1
\end{align}
where we have split the messages into $B$ and $B^*$ components for convenience
and used $y_{k_1:k_2}$ to denote $(y_{k_1},\ldots,y_{k_2})$. $D_{t+1}$
represents the duration of the segment beginning at time $t+1$. The
conditioning on the parameters of the distributions, namely the observation,
duration, and transition parameters, is suppressed from the notation.

We write $F_t = 1$ to indicate a new segment begins at $t+1$ \citep{murphy},
and so to compute the message from $t+1$ to $t$ we sum over all possible
lengths $d$ for the segment beginning at $t+1$, using the backwards message at
$t+d$ to provide aggregate future information given a boundary just after
$t+d$. The final additive term in the expression for $B_t^*(i)$ is described in
\citep{guedon}; it constitutes the contribution of state segments that run off
the end of the provided observations, as per the censoring assumption, and
depends on the survival function of the duration distribution.

Though a very similar
message-passing subroutine is used in HMM Gibbs samplers, there are
significant differences in computational cost between the HMM and HSMM message
computations.  The greater expressive power of the HSMM model necessarily
increases the computational cost: the above
message passing requires $\mathcal{O}(T^2N + TN^2)$ basic operations for a
chain of length $T$ and state cardinality $N$, while the corresponding HMM
message passing algorithm requires only $\mathcal{O}(TN^2)$. However, if the
support of the duration distribution is limited, or if we truncate possible
segment lengths included in the inference messages to some maximum
$d_{\text{max}}$, we can instead express the asymptotic message passing cost as
$\mathcal{O}(T d_{\text{max}} N + TN^2)$. Such truncations are often natural as
the duration prior often causes the message contributions to decay rapidly with
sufficiently large $d$. Though the increased complexity of message-passing over
an HMM significantly increases the cost per iteration of sampling inference for
a global model, the cost is offset because HSMM samplers need far fewer total
iterations to converge.  See the experiments in Section~\ref{sec:experiments}.

\subsection{The HDP-HMM and Sticky HDP-HMM}
The HDP-HMM \citep{originalhdp} provides a natural Bayesian nonparametric
treatment of the classical Hidden Markov Model. The model employs an HDP prior
over an infinite state space, which enables both inference of state complexity
and Bayesian mixing over models of varying complexity.
We provide a brief overview of the HDP-HMM model
and relevant inference algorithms, which we extend to develop the HDP-HSMM\@. A
much more thorough treatment of the HDP-HMM can be found in, for example,
\citep{emilythesis}.

\begin{figure}[tbp] \centering
    \includegraphics[height=2in]{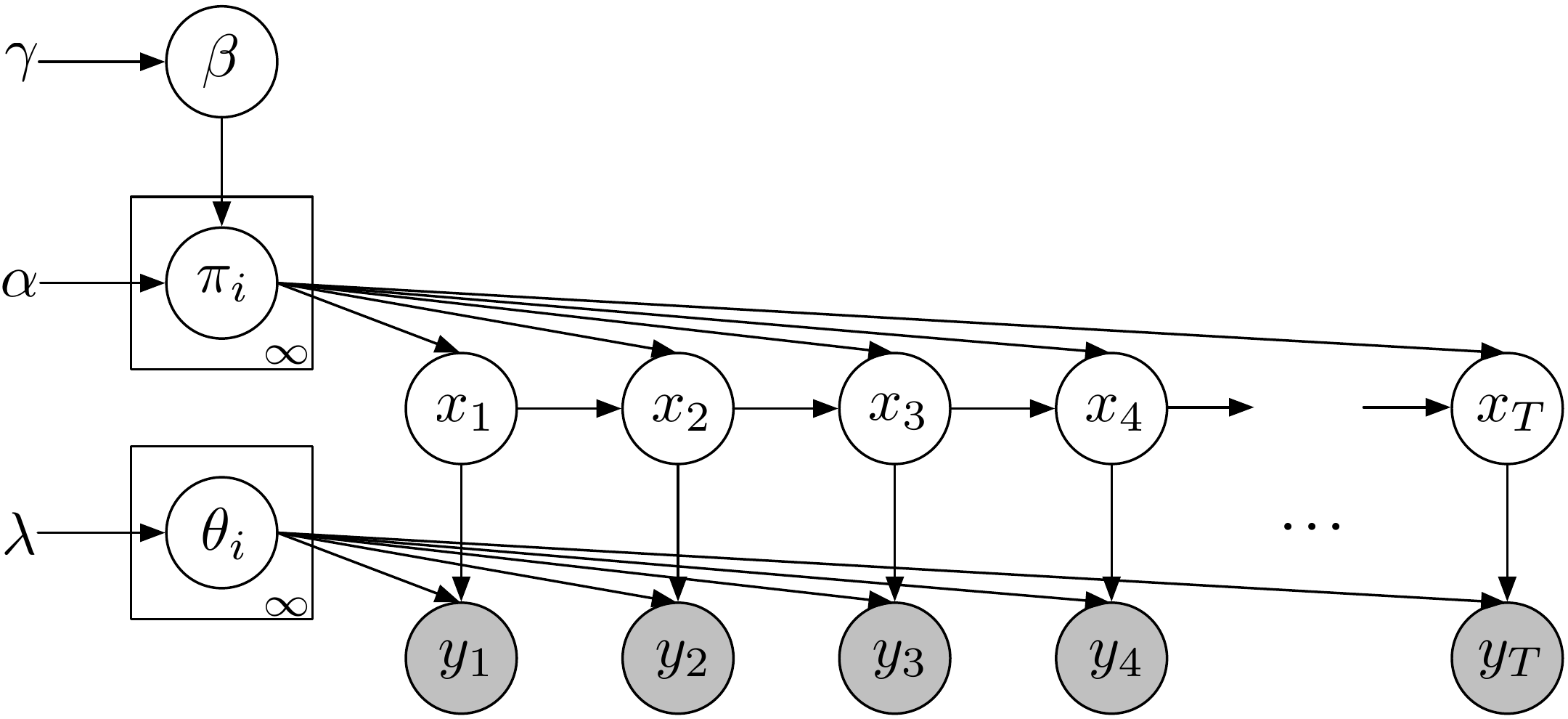}
    \caption{Graphical model for the HDP-HMM.} \label{fig:hdphmm} \end{figure}

The generative process $\text{HDP-HMM}(\gamma,\alpha,H)$ given concentration
parameters $\gamma,\alpha > 0$ and base measure (observation prior) $H$ can be
summarized as:
\begin{align}
    \beta & \sim \text{GEM}(\gamma) & & & \\
    \pi_i & \iid{\sim} \text{DP}(\alpha,\beta) &
        \theta_i \iid{\sim} H & & i=1,2,\ldots \\
    x_t &\sim \pi_{x_{t-1}} & & & \\
    y_t &\sim f(\theta_{x_t}) & & & t=1,2,\ldots,T
\end{align}
where GEM denotes a stick breaking process \citep{GEM} and $f$ denotes an
observation distribution parameterized by draws from $H$. We set $x_1 := 1$. We
have also suppressed explicit conditioning from the notation.  See
Figure~\ref{fig:hdphmm} for a graphical model.

The HDP plays the role of a prior over infinite transition matrices: each
$\pi_j$ is a DP draw and is interpreted as the transition distribution from
state $j$, i.e.~the $j$th row of the infinite transition matrix. The $\pi_j$
are linked by being DP draws parameterized by the same discrete measure
$\beta$, thus $\mathbb{E}[\pi_j] = \beta$ and the transition distributions tend
to have their mass concentrated around a typical set of states, providing the
desired bias towards re-entering and re-using a consistent set of states.

The Chinese Restaurant Franchise and direct-assignment collapsed sampling methods described in
\citep{originalhdp, emilythesis} are approximate
inference algorithms for the full infinite dimensional HDP, but they have a particular
weakness in the sequential-data context of the HDP-HMM: each state transition
must be re-sampled individually, and strong correlations within the state
sequence significantly reduce mixing rates\citep{emilythesis}.  As a result, finite approximations to the HDP have been
studied for the purpose of providing faster mixing.
Of particular note is the popular weak limit approximation, used in
\citep{emilysticky}, which has been shown to reduce mixing times for HDP-HMM
inference while sacrificing little of the ``tail'' of the infinite transition
matrix. In this paper, we describe how the HDP-HSMM with geometric durations
can provide an HDP-HMM sampling inference algorithm that maintains the ``full''
infinite-dimensional sampling process while mitigating the detrimental mixing
effects due to the strong correlations in the state sequence, thus providing a
new alternative to existing HDP-HMM sampling methods.

The Sticky HDP-HMM augments the HDP-HMM with an extra parameter $\kappa>0$ that
biases the process towards self-transitions and thus provides a method to
encourage longer state durations. The
$\text{Sticky-HDP-HMM}(\gamma,\alpha,\kappa,H)$ generative process can be
written
\begin{align}
    \beta & \sim \text{GEM}(\gamma) & & & \\
    \pi_i & \iid{\sim} \text{DP}(\alpha + \kappa,\beta + \kappa \delta_{j}) &
        \theta_i  \iid{\sim} H & & i=1,2,\ldots \\
    x_t &\sim \pi_{x_{t-1}} & & & \\
    y_t &\sim f(\theta_{x_t}) & & & t=1,2,\ldots,T
\end{align}
where $\delta_j$ denotes an indicator function that takes value $1$ at index
$j$ and $0$ elsewhere.  While the Sticky HDP-HMM allows some control over
duration statistics, the state duration distributions remain geometric; the
goal of this work is to provide a model in which any duration distributions may
be used.

\section{Models}
\label{sec:models}
In this section, we introduce the explicit-duration HSMM-based models that we
use in the remainder of the paper. We define the finite Bayesian HSMM and
the HDP-HSMM and show how they can be used as components in more complex
models, such as in a factorial structure.
We describe generative processes that do not allow self-transitions in the
state sequence, but we emphasize that we can also allow self-transitions and
still employ the inference algorithms we describe; in fact, allowing
self-transitions simplifies inference in the HDP-HSMM, since complications
arise as a result of the hierarchical prior and an elimination of
self-transitions. However, there is a clear modeling gain by eliminating
self-transitions: when self-transitions are allowed, the ``explicit duration
distributions'' do not model the state duration statistics directly. To allow
direct modeling of state durations, we must consider the case where
self-transitions do not occurr.

We do not investigate here the problem of selecting particular observation and
duration distribution classes; model selection is a fundamental challenge in
generative modeling, and models must be chosen to capture structure in any
particular data. Instead, we provide the HDP-HSMM and related models as tools
in which modeling choices (such as the selection of observation and duration
distribution classes to fit particular data) can be made flexibly and
naturally.

\subsection{Finite Bayesian HSMM}
The finite Bayesian HSMM is a combination of the Bayesian HMM approach with
semi-Markov state durations and is the model we generalize to the
HDP-HSMM\@. Some forms of finite Bayesian HSMMs have been described previously,
such as in \citep{hashimoto2009bayesian} which treats observation parameters as
Bayesian latent variables, but to the best of our knowledge the first fully
Bayesian treatment of all latent components of the HSMM was given in
\citep{johnson2010hdphsmm} and later independently in
\citep{dewar2012inference}, which allows self-transitions.

It is instructive to compare this construction with that of the finite model
used in the weak-limit HDP-HSMM sampler that will be described in
Section~\ref{sec:weaklimit}, since in that case the hierarchical ties between
rows of the transition matrix requires particular care.

The generative process for a Bayesian HSMM with $N$ states and observation and
duration parameter prior distributions of $H$ and $G$, respectively, can be
summarized as

\begin{align}
    \pi_i & \iid{\sim} \text{Dir}(\alpha (1- \delta_i)) &
        (\theta_i, \omega_i) \iid{\sim} H \times G & & i=1,2,\ldots,N \\
    z_s &\sim \pi_{z_{s-1}} & & & \\
    D_s &\sim g(\omega_{z_s}) & & & s=1,2,\ldots \\
    x_{t_s^1:t_s^2} &= z_s & & & \\
    y_{t_s^1:t_s^2} &\iid{\sim} f(\theta_{z_s}) & &
    t_s^1=\sum_{\bar{s}<s} D_{\bar{s}}  & t_s^2 = t_s^1 + D_s - 1
\end{align}
where $f$ and $g$ denote observation and duration distributions parameterized
by draws from $H$ and $G$, respectively. The indices $t_s^1$ and $t_s^2$ denote
the first and last index of segment $s$, respectively, and $x_{t_s^1:t_s^2} :=
(x_{t_s^1},x_{t_s^1+1},\ldots,x_{t_s^2})$. We use $\text{Dir}(\alpha (1-\delta_i))$
to denote a symmetric Dirichlet distribution with parameter $\alpha$ except
with the $i$th component of the hyperparameter vector set to zero, hence fixing
$\pi_{ii}=0$ and ensuring there will be no self-transitions sampled in the
\emph{super-state sequence} $(z_s)$. We also define the \emph{label sequence}
$(x_t)$ for convenience; the pair $(z_s,D_s)$ is the run-length encoding of
$(x_t)$. The process as written generates an infinite sequence of observations; we
observe a finite prefix of size $T$.

Note, crucially, that in this definition the $\pi_i$ are not tied across
various $i$. In the HDP-HSMM, as well as the weak limit model used for
approximate inference in the HDP-HSMM, the $\pi_i$ will be tied through the
hierarchical prior (specifically via $\beta$), and that connection is necessary
to penalize the total number of states and encourage a small, consistent set of
states to be visited in the state sequence. However, the interaction between
the hierarchical prior and the elimination of self-transitions present an
inference challenge.

\subsection{HDP-HSMM}

\begin{figure}[tbp]
    \centering
    \includegraphics[width=3.5in]{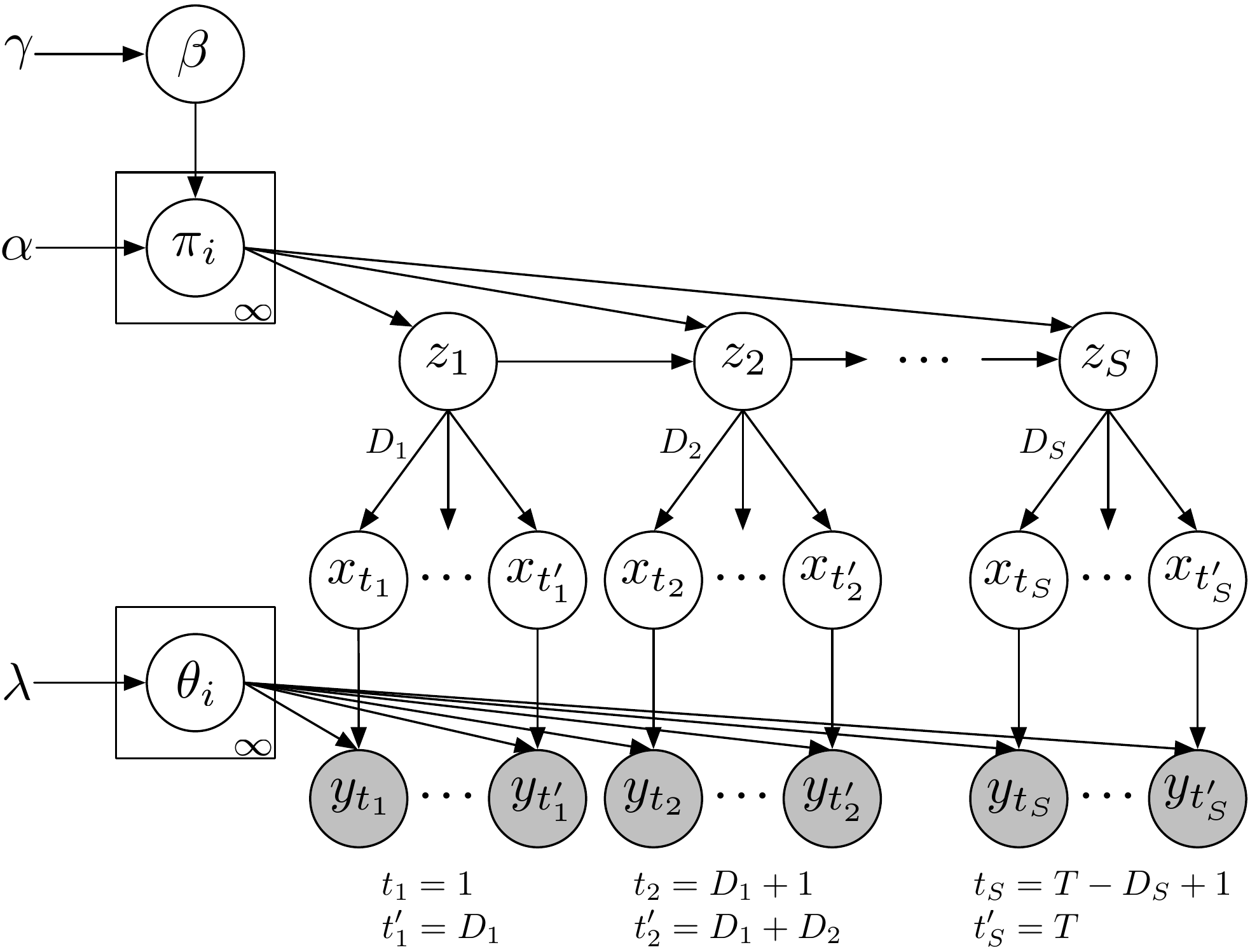}
    \caption{A graphical model for the HDP-HSMM in which the number of segments
        $S$, and hence the number of nodes, is random.}
    \label{fig:hdphsmm_fullgraph}
\end{figure}

The generative process of the HDP-HSMM is similar to that of the HDP-HMM (as
described in, e.g., \citep{emilysticky}), with some extra work to include
duration distributions. The process $\text{HDP-HSMM}(\gamma,\alpha,H,G)$,
illustrated in Figure~\ref{fig:hdphsmm_fullgraph}, can be written
\begin{align}
    \label{eq:hdphsmmstart}
    \beta & \sim \text{GEM}(\gamma) & & & \\
    \pi_i & \iid{\sim} \text{DP}(\alpha,\beta) &
        (\theta_i,\omega_i) \iid{\sim} H \times G & & i=1,2,\ldots \\
    z_s &\sim \bar{\pi}_{z_{s-1}} & & & \\
    D_s &\sim g(\omega_{z_s}) & & & s=1,2,\ldots \\
    x_{t_s^1:t_s^2} &= z_s & & & \\
    y_{t_s^1:t_s^2} &\iid{\sim} f(\theta_{x_t}) & &
        t_s^1=\sum_{\bar{s}<s} D_{\bar{s}}  & t_s^2 = t_s^1 + D_s - 1
    \label{eq:hdphsmmend}
\end{align}
where we use $\bar{\pi}_i := \frac{\pi_{ij}}{1-\pi_{ii}} ( 1 - \delta_{ij} )$ to eliminate
self-transitions in the \emph{super-state sequence} $(z_s)$. As with the finite
HSMM, we define the \emph{label sequence} $(x_t)$ for convenience. We observe a
finite prefix of size $T$ of the observation sequence.

Note that the atoms we edit to eliminate self-transitions are the same atoms
that are affected by the global sticky bias in the Sticky HDP-HMM.

\subsection{Factorial Structure}
We can easily compose our sequential models into other common model structures,
such as the factorial structure of the factorial HMM
\citep{ghahramani1997factorial}. Factorial models are very useful for source
separation problems, and when combined with the rich class of sequential models
provided by the HSMM, one can use prior duration information about each source
to greatly improve performance (as demonstrated in
Section~\ref{sec:experiments}).  Here, we briefly outline the factorial model
and its uses.

If we use $y \sim \text{HDP-HSMM}(\alpha,\gamma,H,G)$ to denote an observation
sequence generated by the process defined in
\Crefrange{eq:hdphsmmstart}{eq:hdphsmmend}, then the generative process for a
factorial HDP-HSMM with $K$ component sequences can be written
\begin{align}
    y^{(k)} \sim \text{HDP-HSMM}(\alpha_k,\gamma_k,H_k,G_k) & & k=1,2,\ldots,K \\
    \bar{y}_t := \sum_{k=1}^K y_t^{(k)} + w_t & & t=1,2,\ldots,T
\end{align}
where $w_t$ is a noise process independent of the other components of the
model states.

\begin{figure}[pt]
    \centering
    \includegraphics[width=2.5in]{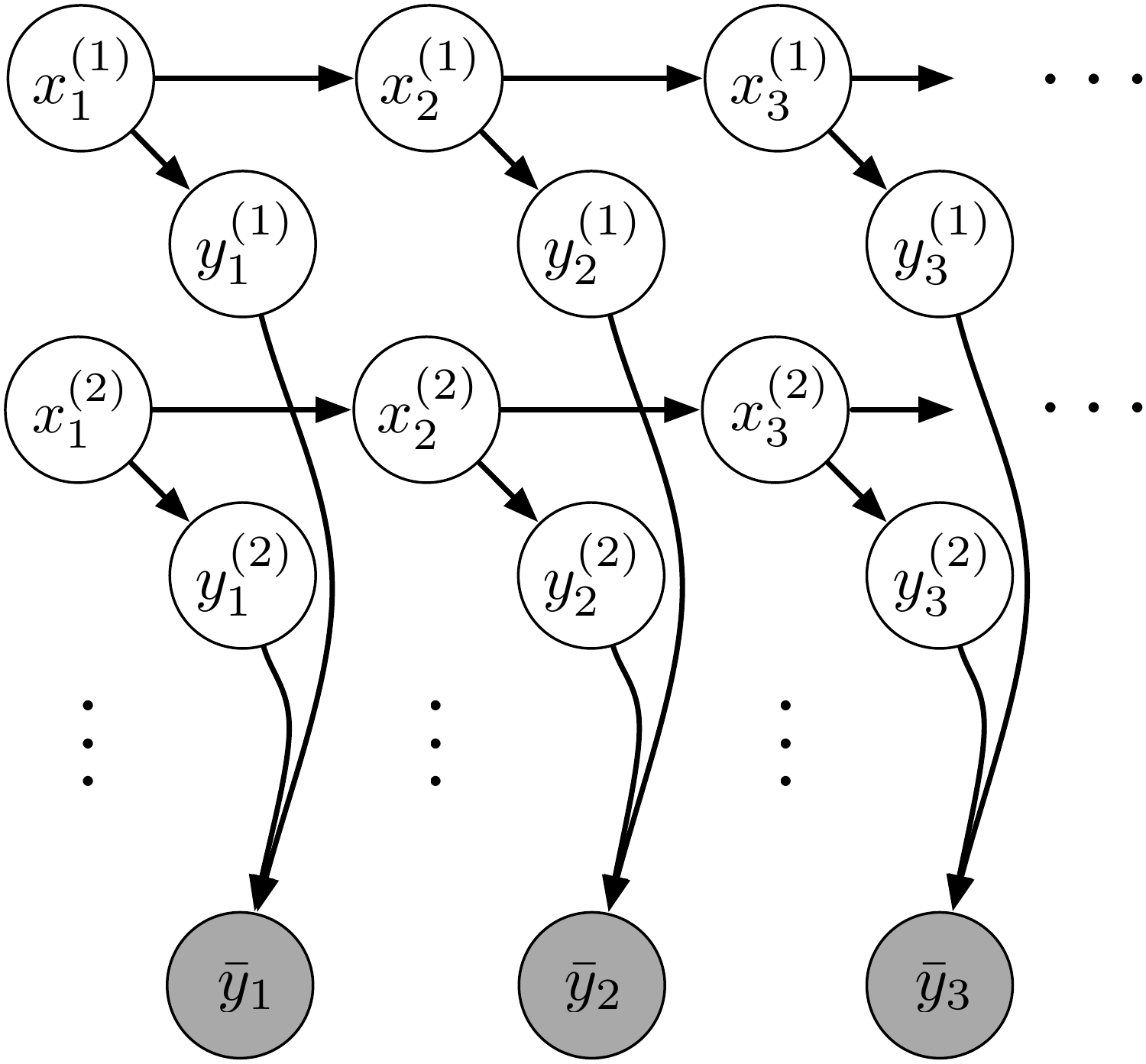}
    \caption{A graphical model for the factorial HMM, which can naturally be
        extended to factorial structures involving the HSMM or HDP-HSMM.}
    \label{fig:factorial_hmm}
\end{figure}

A graphical model for a factorial HMM can be seen in
Figure~\ref{fig:factorial_hmm}, and a factorial HSMM or factorial HDP-HSMM
simply replaces the hidden state chains with semi-Markov chains. Each chain,
indexed by superscripts, evolves with independent dynamics and produces
independent emissions, but the observations are combinations of the independent
emissions. Note that each component HSMM is not restricted to any fixed number
of states.

Such factorial models are natural ways to frame source separation or
disaggregation problems, i.e., identifying component emissions and component
states. With the Bayesian framework, we also model uncertainty and ambiguity in
such a separation. In Section~\ref{sec:power-disaggregation} we demonstrate the
use of a factorial HDP-HSMM for the task of disaggregating home power signals.

Problems in source separation or disaggregation are often ill-conditioned, and
so one relies on prior information in addition to the source independence
structure to solve the separation problem. Furthermore, representation of
uncertainty is often important, since there may be several good explanations
for the data. These considerations motivate Bayesian inference as well as
direct modeling of state duration statistics.

\section{Inference Algorithms}
\label{sec:algorithms}
We describe two Gibbs sampling inference algorithms, beginning with a
sampling algorithm for the finite Bayesian HSMM, which is built upon in
developing algorithms for the HDP-HSMM in the sequel. Next, we develop a
weak-limit Gibbs sampling algorithm for the HDP-HSMM, which parallels the
popular weak-limit sampler for the HDP-HMM and its sticky extension.  Finally,
we introduce a collapsed sampler which parallels the direct assignment
sampler of \citep{originalhdp}.
%For a summary of the HDP-HSMM samplers, see Table~\ref{tab:hdp-hsmm-samplers}.
For all both of the HDP-HSMM samplers there is a loss of conjugacy with the
HDP prior due to the fact that self-transitions in the super-state sequence are
not permitted (see Section~\ref{sec:finite_conjugacy}). We develop auxiliary
variables to form an augmented representation that effectively recovers
conjugacy and hence enables fast exact Gibbs steps.

In comparing the weak limit and direct assignment sampler, the most important
trade-offs are that the direct assignment sampler works with the infinite model
by integrating out the transition matrix $\pi$ while simplifying bookkeeping by
maintaining part of $\beta$; it also collapses the observation and duration
parameters. However, the variables in the label sequence $(x_t)$ are coupled by
the integration, and hence each element of the label sequence must be resampled
sequentially. In contrast, the weak limit sampler represents all latent
components of the model (up to an adjustable finite approximation for the HDP)
and thus allows block resampling of the label sequence by exploiting HSMM
message passing.
% The second direct assignment sampler we develop for the HDP-HSMM, the
% super-state direct assignment sampler, has the advantage of being able to
% re-sample entire segment labels in each step, thus avoiding the main cause of
% slow mixing in the HDP-HMM direct assignment sampler and hence also offering an
% alternative HDP-HMM sampling method.

We end the section with a discussion of leveraging changepoint side-information
to greatly accelerate inference.

% \begin{table}[htb]
%     \centering
%     \begin{tabular}{|c|p{0.15\textwidth}|p{0.45\textwidth}|}
%         \hline
%         Name & Represents & Advantages / Disadvantages \\\hline
%         \hline
%         Weak Limit & $\{\pi_i\},\{\theta_i\},$\newline$\{\omega_i\},(x_t)$ & -
%         block-resample state sequence \newline - must represent and resample all
%         parameters \\\hline

%         Label Sequence DA Sampler & $\beta,(x_t)$ & - collapses observation and
%         duration parameters as well as most of the HDP \newline - slow to mix because
%         labels $x_t$ must be resampled one at a time \\\hline

%         Super-State DA Sampler & $\beta,(z_s),(D_s)$ & - also collapses
%         observation and duration parameters as well as most of the HDP \newline -
%         re-assigns state labels for entire segments in each step \newline - slow at
%         mixing over number of segments \\\hline
%     \end{tabular}
%     \caption{Table summary of HDP-HSMM samplers. DA stands for direct
%         assignment. The second column lists the representation that is
%         maintained by the sampler (i.e.~the latent variables that are
%         explicitly resampled in Gibbs steps). Auxiliary variables are not
%         listed.}
%     \label{tab:hdp-hsmm-samplers}
% \end{table}

\subsection{A Gibbs Sampler for the Finite Bayesian HSMM}
\label{sec:finite_hsmm_gibbs}

\subsubsection{Outline of Gibbs Sampler}
To perform posterior inference in a finite Bayesian HSMM,
we construct a Gibbs sampler resembling that for finite HMMs.
Our goal is to construct samples from the posterior
\begin{align}
    p((x_t), \{\theta_i\}, \{\pi_i\}, \{\omega_i\} | (y_t), H, G, \alpha)
\end{align}
by drawing samples from the distribution, where $G$ represents the prior
over duration parameters. We can construct these samples by following a Gibbs
sampling algorithm in which we iteratively sample from the appropriate conditional distributions of $(x_t)$, $\{\pi_i\}$, $\{\omega_i\}$, and $\{\theta_i\}$.

Sampling $\{\theta_i\}$ or $\{\omega_i\}$ from their respective conditional
distributions can be easily reduced to standard problems depending on the
particular priors chosen.  Sampling the transition matrix rows $\{\pi_i\}$ is
straightforward if the prior on each row is Dirichlet over the off-diagonal
entries and so we do not discuss it in this section, but we note that when the
rows are tied together hierarchically (as in the weak-limit approximation to
the HDP-HSMM), resampling the $\{\pi_i\}$ correctly requires particular care
(see Section~\ref{sec:finite_conjugacy}).

Sampling $(x_t) | \{\theta_i\}, \{\pi_i\}, (y_t)$ in a finite Bayesian Hidden
semi-Markov Model was first introduced in \citep{johnson2010hdphsmm} and, in
independent work, later in \cite{dewar2012inference}. In the following section we
develop the algorithm for block-sampling the state sequence $(x_t)$ from its
conditional distribution by employing the HSMM message-passing scheme.

\subsubsection{Blocked Conditional Sampling of $(x_t)$ with Message Passing}
\label{sec:finite-hsmm-statesampling}

To block sample $(x_t) | \{\theta_i\}, \{\pi_i\}, \{\omega_i\}, (y_t)$ in an
HSMM we can extend the standard block state sampling scheme for an HMM\@. The
key challenge is that to block sample the states in an HSMM we must also be
able to sample the posterior duration variables.

If we compute the backwards messages $B$ and $B^*$ described in
Section~\ref{sec:classical_hsmm}, then we can easily draw a posterior sample
for the first state according to:
\begin{align}
    p(x_1 = k | y_{1:T}) &\propto p(x_1 = k) p(y_{1:T} | x_1 = k,F_0=1) \\
    &= p(x_1 = k) B^*_0(k)
\end{align}
where we have used the assumption that the observation sequence begins on a
segment boundary ($F_0=1$) and suppressed notation for conditioning on
parameters.

We can also use the messages to efficiently draw a sample from the posterior
duration distribution for the sampled initial state. Conditioning on the
initial state draw, $\bar{x}_1$, the posterior duration of the first state is:
\begin{align}
    p(&D_1 = d | y_{1:T}, x_1 = \bar{x}_1, F_0 = 1) = \frac{p(D_1=d,y_{1:T} | x_1 = \bar{x}_1, F_0)}{p(y_{1:T} | x_1 = \bar{x}_1, F_0)} \\
    &= \frac{p(D_1 = d | x_1=\bar{x}_1, F_0) p(y_{1:d} | D_1 = d, x_1 = \bar{x}_1, F_0) p(y_{d+1:T} | D_1 = d, x_1 = \bar{x}_1, F_0) }{p(y_{1:T} | x_1=\bar{x}_1, F_0 )}\\
    &= \frac{p(D_1=d) p(y_{1:d} | D_1 = d, x_1 = \bar{x}_1, F_0 = 1) B_d(\bar{x}_1)}{B_0^*(\bar{x}_1)}.
\end{align}
We repeat the process by using $x_{D_1+1}$ as our new initial
state with initial distribution $p(x_{D_1+1}=i | x_1 = \bar{x}_1)$,
and thus draw a block sample for the entire label sequence.

\subsection{A Weak-Limit Gibbs Sampler for the HDP-HSMM}
\label{sec:weaklimit}
The weak-limit sampler for an HDP-HMM \citep{emilysticky} constructs a finite
approximation to the HDP transitions prior with finite $L$-dimensional
Dirichlet distributions, motivated by the fact that the infinite limit of such
a construction converges in distribution to a true HDP:
\begin{align}
    \beta | \gamma &\sim \text{Dir}(\gamma/L,\ldots,\gamma/L) &~\\
    \pi_i | \alpha, \beta &\sim \text{Dir}(\alpha \beta_1,\ldots,\alpha \beta_L) & i=1,\ldots,L
\end{align}
where we again interpret $\pi_i$ as the transition distribution for state $i$
and $\beta$ as the distribution which ties state distributions together and
encourages shared sparsity. Practically, the weak limit approximation enables
the complete representation of the transition matrix in a finite form, and thus,
when we also represent all parameters, allows block sampling of the entire
label sequence at once, resulting in greatly accelerated mixing in many
circumstances. The parameter $L$ gives us control over the approximation, with
the guarantee that the approximation will become exact as $L$ grows; see
\citep{ishwaran2000markov}, especially Theorem 1, for a discussion of
theoretical guarantees. Note that the weak limit approximation is more
convenient for us than the truncated stick-breaking approximation because it
directly models the state transition probabilities, while stick lengths in the
HDP do not directly represent state transition probabilities because multiple
sticks in constructing $\pi_i$ can be sampled at the same atom of $\beta$.

We can employ the weak limit approximation to create a finite
HSMM that approximates inference in the HDP-HSMM\@.
This approximation technique often results in greatly accelerated mixing,
and hence it is the technique we employ for the experiments in the sequel.
However, the inference algorithm of Section~\ref{sec:finite_hsmm_gibbs} must be
modified to incorporate the fact that the $\{\pi_i\}$ are no longer mutually
independent and are instead tied through the shared $\beta$. This dependence
between the transition rows introduces potential conjugacy issues with the
hierarchical Dirichlet prior; the following section explains the difficulty as
well as a clean solution via auxiliary variables.

The beam sampling technique \cite{van2008beam} can be applied here with little
modification, as in \cite{dewar2012inference}, to sample over the approximation
parameter $L$, thus avoiding the need to set $L$ a priori while still allowing
instantiation of the transition matrix and block sampling of the state
sequence. This technique is especially useful if the number of states could be
very large and is difficult to bound a priori. We do not explore beam sampling
here.

\subsubsection{Conditional Sampling of $\{\pi_i\}$ with Data Augmentation}
\label{sec:finite_conjugacy}

To construct our overall Gibbs sampler, we need to be able to easily resample
the transition matrix $\pi$ given the other components of the model.
However, by ruling out self-transitions while maintaining a hierarchical link
between the transition rows, the model is no longer fully conjugate, and hence
resampling is not necessarily easy.  To observe the loss of conjugacy using the
hierarchical prior required in the weak-limit approximation, note that we can
summarize the relevant portion of the generative model as
\begin{align}
    \beta | \gamma &\sim \text{Dir}(\gamma, \ldots, \gamma)\\
    \pi_j | \beta &\sim \text{Dir}(\alpha \beta_1,\ldots,\alpha \beta_L) & j=1,\ldots,L\\
    x_t | \{\pi_j\},x_{t-1} &\sim \bar{\pi}_{x_{t-1}} & t=2,\ldots,T\\
\end{align}
where $\bar{\pi}_j$ represents $\pi_j$ with the $j$th component removed and
renormalized appropriately:
\begin{align}
    \bar{\pi}_{ji} = \frac{\pi_{ji}(1- \delta_{ij})}{1 - \pi_{jj}}
\end{align}
with $\delta_{ij} = 1$ if $i=j$ and $\delta_{ij}=0$ otherwise. The
deterministic transformation from $\pi_j$ to $\bar{\pi}_j$ eliminates
self-transitions. Note that we have suppressed the observation parameter set,
duration parameter set, and observation sequence sampling for simplicity.

Consider the distribution of $\pi_1 | (x_t),\beta$, the first row of the
transition matrix:
\begin{align}
    p(\pi_1 | (x_t), \beta) &\propto p(\pi_1 | \beta) p((x_t) | \pi_1) \\
    &\propto \pi_{11}^{\alpha\beta_1 - 1} \pi_{12}^{\alpha\beta_2 - 1} \cdots \pi_{1L}^{\alpha\beta_L - 1} \left( \frac{\pi_{12}}{1 - \pi_{11}} \right)^{n_{12}} \left( \frac{\pi_{13}}{1 - \pi_{11}} \right)^{n_{13}} \cdots \left( \frac{\pi_{1L}}{1 - \pi_{11}} \right)^{n_{1L}} 
\end{align}
where $n_{ij}$ are the number of transitions from state $i$ to state $j$ in the
state sequence $(x_t)$. Essentially, because of the extra
$\frac{1}{1-\pi_{11}}$ terms from the likelihood without self-transitions, we
cannot reduce this expression to the Dirichlet form over the components of
$\pi_{1}$, and therefore we cannot proceed with sampling $m$ and resampling
$\beta$ and $\pi$ as in \cite{originalhdp}.

However, we can introduce auxiliary variables to recover conjugacy, following
the general data augmentation technique described in \citep{van2001art}. We define an
extended generative model with extra random variables, and then show through
simple manipulations that conditional distributions simplify with the
additional variables, hence allowing us to cycle simple Gibbs updates to
produce a sampler.

\begin{figure}[tp]
    \centering
    \subfigure[]{\includegraphics[scale=0.5]{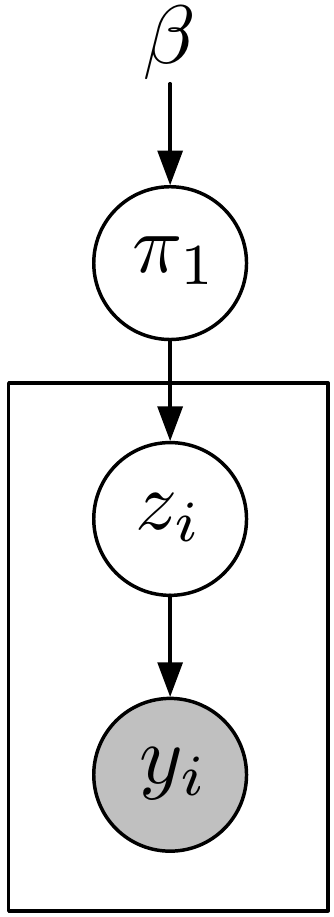} \label{fig:simpleconjugacy:plain}}
    \subfigure[]{\includegraphics[scale=0.5]{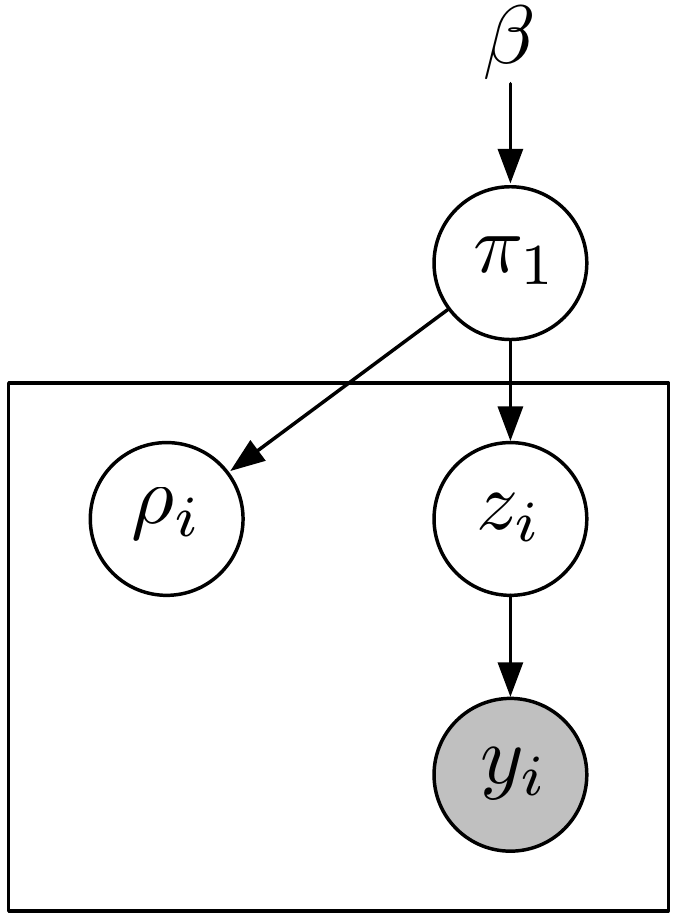} \label{fig:simpleconjugacy:aux}}
    \caption{Simplified depiction of the relationship between the auxiliary variables and the rest of the model;~\ref{fig:simpleconjugacy:plain} depicts the nonconjugate setting and~\ref{fig:simpleconjugacy:aux} shows the introduced auxiliary variables $\{\rho_i\}$.}
    \label{fig:simpleconjugacy}
\end{figure}

For simplicity, we focus on the first row of the transition matrix,
namely $\pi_1$, and the draws that depend on it; the reasoning easily extends
to the other rows. We also drop the parameter $\alpha$ for convenience. First,
we write the relevant portion of the generative process as
\begin{align}
    \pi_1 | \beta &\sim \text{Dir}(\beta)\\
    z_i | \bar{\pi}_1 &\sim \bar{\pi}_1 \quad i=1,\ldots,n\\
    y_i | z_i &\sim f(z_i) \quad i=1,\ldots,n
\end{align}
Here, $n$ counts the total number of transitions out of state $1$ and the
$\{z_i\}$ represent the transitions out of state $1$ to a specific state:
sampling $z_i=k$ represents a transition from state $1$ to state $k$. The
$\{y_i\}$ represent the observations on which we condition; in particular,
if we have $z_i=k$ then the $y_i$ corresponds to an emission from state $k$ in
the HSMM\@. See the graphical model in Figure~\ref{fig:simpleconjugacy:plain}
for a depiction of the relationship between the variables.

We can introduce auxiliary variables $\{\rho_i\}_{i=1}^n$, where each $\rho_i$
is independently drawn from a geometric distribution supported on
$\{0,1,\ldots\}$ with success parameter $1-\pi_{11}$, i.e. $\rho_i | \pi_{11}
\sim \text{Geo}(1-\pi_{11})$ (See Figure~\ref{fig:simpleconjugacy:aux}). Thus
our posterior becomes:
\begin{align}
    p(\pi_1 | \{z_i\}, \{\rho_i\}) &\propto p(\pi_1) p(\{z_i\} | \pi_1) p(\{\rho_i\} | \{\pi_{1i}\})\\
    &\propto \pi_{11}^{\beta_1 - 1} \pi_{12}^{\beta_2 - 1} \cdots \pi_{1L}^{\beta_L - 1} \left( \frac{\pi_{12}}{1 - \pi_{11}} \right)^{n_{2}} 
    \cdots \left( \frac{\pi_{1L}}{1 - \pi_{11}} \right)^{n_{L}} \left(\prod_{i=1}^n \pi_{11}^{\rho_i} (1-\pi_{11}) \right)\\
    &= \pi_{11}^{\beta_1 + \sum_i \rho_i - 1} \pi_{12}^{\beta_2 + n_2 - 1} \cdots \pi_{1L}^{\beta_L + n_L - 1}\\
    &\propto \text{Dir}\left(\beta_1 + \sum_i \rho_i, \beta_2 + n_2, \ldots, \beta_L + n_L \right).
\end{align}
Noting that $n = \sum_i n_i$, we recover conjugacy and hence can iterate simple
Gibbs steps.

\begin{figure}
    \centering
    \includegraphics[width=3in]{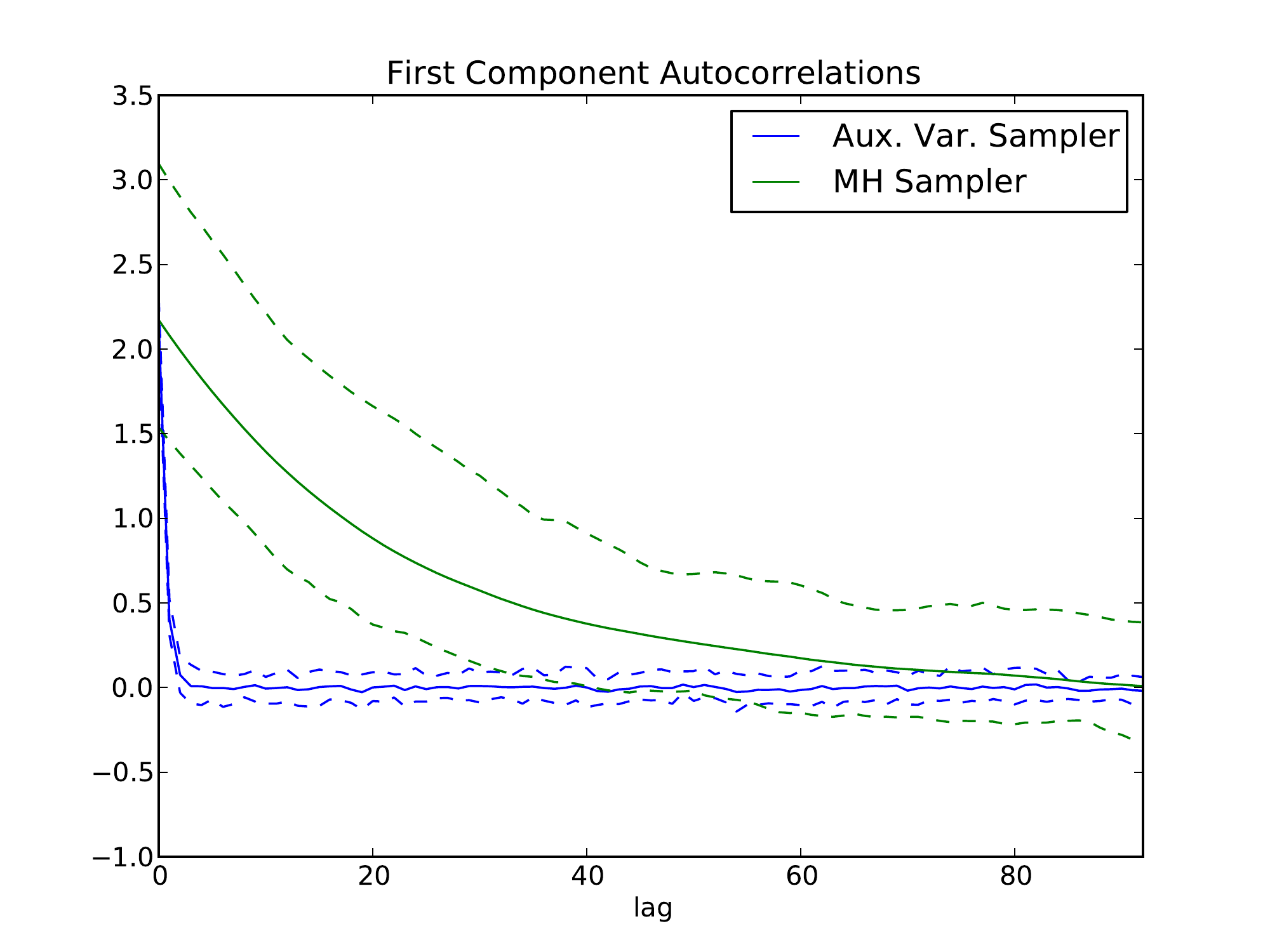}
    \caption{Empirical sample chain autocorrelation for the first component of
        $\pi$ for both the proposed auxiliary variable sampler and a
        Metropolis-Hastings sampler. The rapidly diminishing autocorrelation
        for the auxiliary variable sampler is indicative of fast mixing.}
    \label{fig:aux_autocorr}
\end{figure}

\begin{figure}
    \centering
    \subfigure[]{\includegraphics[width=2.5in]{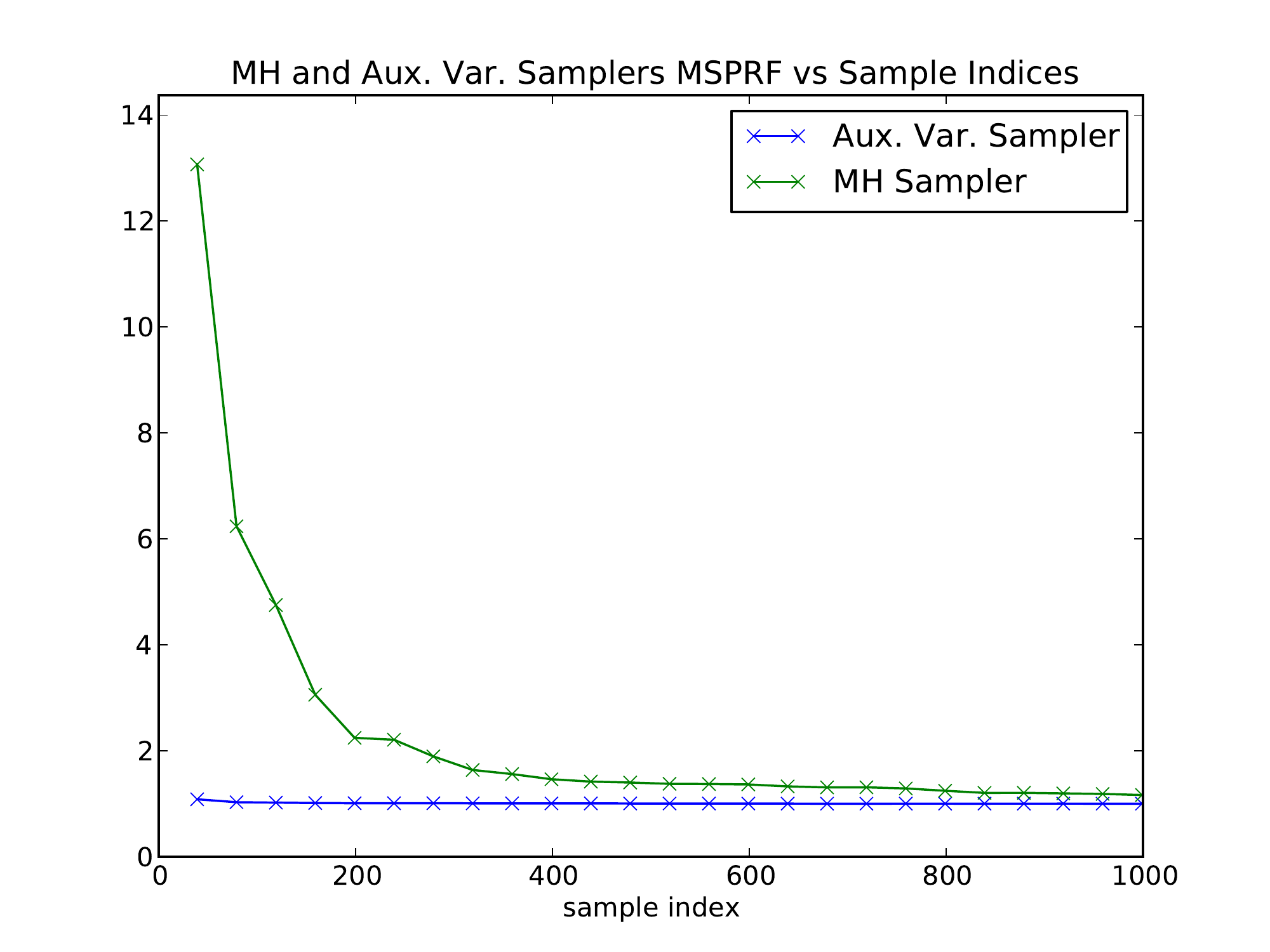}
        \label{fig:aux_mpsrf_sampleindex}}
    \subfigure[]{\includegraphics[width=2.5in]{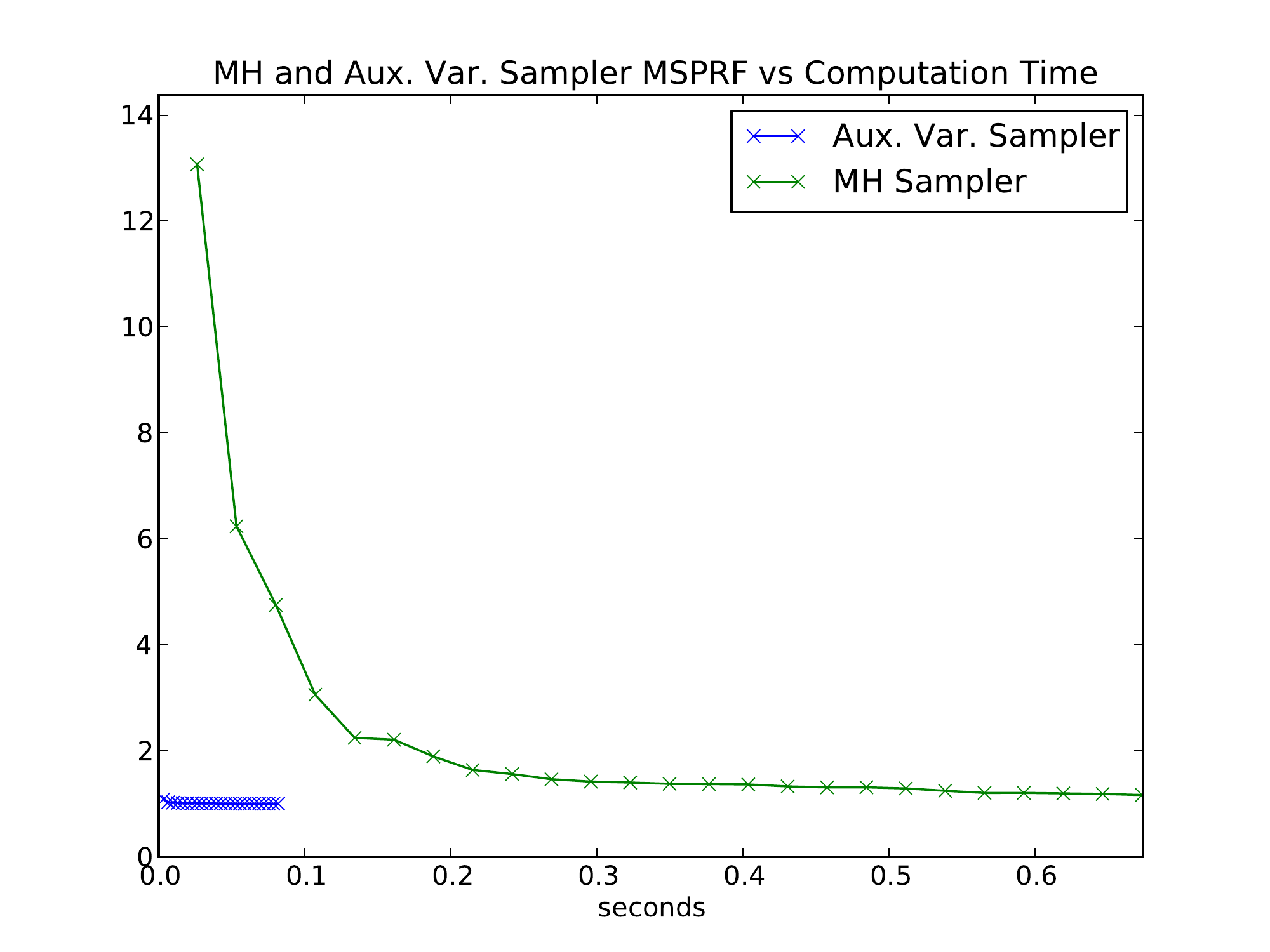}
        \label{fig:aux_mpsrf_timeindex}}
    \caption{Multivariate Potential Scale Reduction Factors for both the
        proposed auxiliary variable sampler and a Metropolis-Hastings sampler.
        The auxiliary variable sampler rapidly achieves the statistic's
        asymptotic value of unity. Note that the auxiliary variable sampler is
        also much more efficient to execute, as shown
        in~\ref{fig:aux_mpsrf_timeindex}.}
    \label{fig:aux_mpsrf}
\end{figure}

We can compare the numerical performance of the auxiliary variable sampler to a
Metropolis-Hastings sampler in the simplified model. For a detailed
evaluation, see \cite{johnson2012truncated}; in deference to space
considerations, we only reproduce two figures from that report here.
Figure~\ref{fig:aux_autocorr} shows the sample chain autocorrelations for the
first component of $\pi$ in both samplers. Figure~\ref{fig:aux_mpsrf} compares
the Multivariate Scale Reduction Factors of \cite{brooks1998general} for the two samplers, where good
mixing is indicated by achieving the statistic's asymptotic value of unity.

% Intuitively, we are able to fill in the data to include self-transitions
% because before each transition is sampled, we must sample and reject
% $\text{Geo}(1-\pi_{11})$ self-transitions. Note that this procedure depends on the
% fact that, by construction, the $\{\rho_i\}$ are conditionally independent of
% the data, $\{y_i\}$, given $\{z_i\}$.

\begin{figure}[tp]
    \centering
    \includegraphics[height=2.5in]{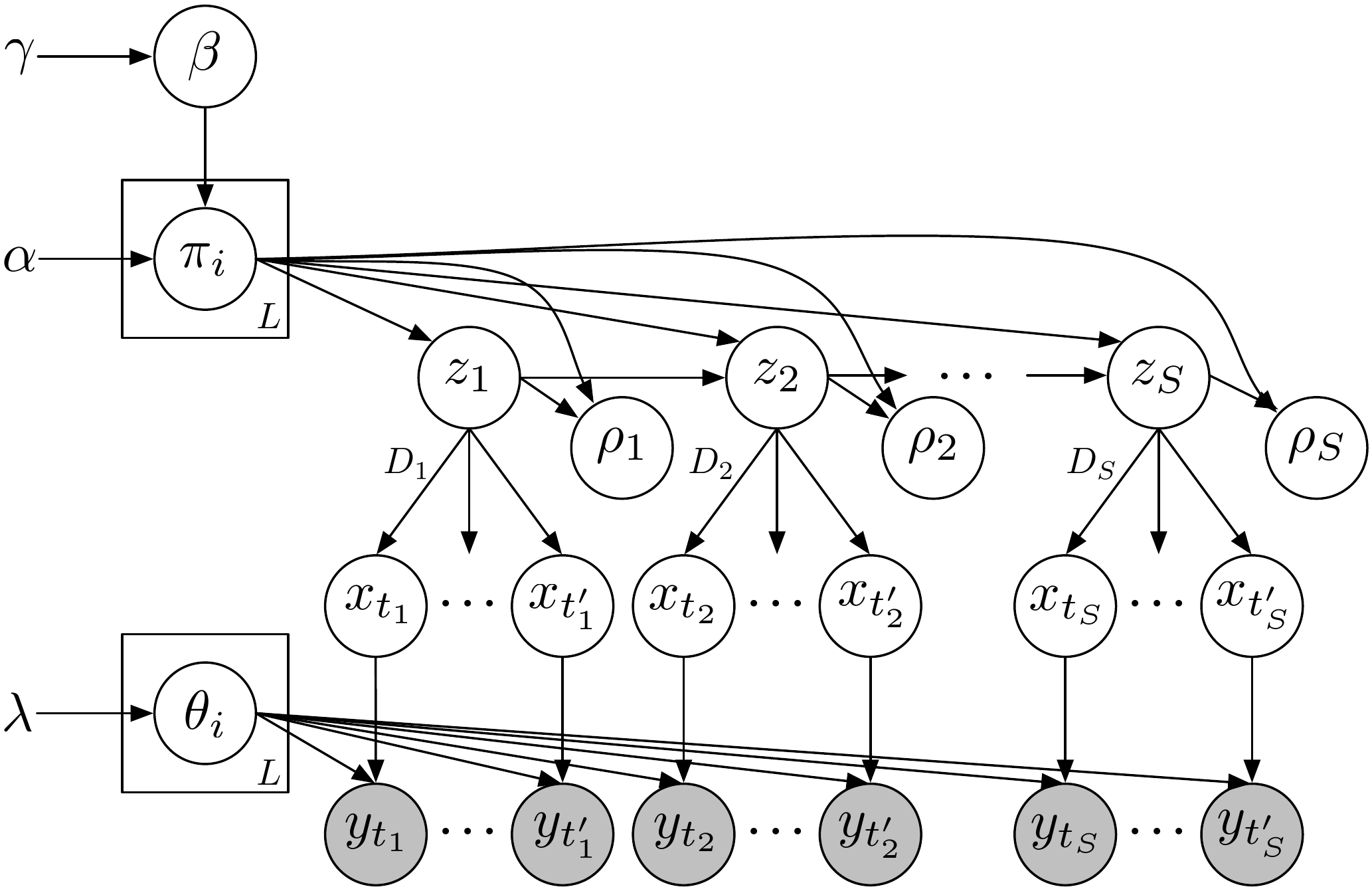}
    \caption{Graphical model for the weak-limit approximation including auxiliary variables. }
    \label{fig:weaklimitfull}
\end{figure}

We can easily extend the data augmentation to the full HSMM, and once we have
augmented the data with the auxiliary variables $\{\rho_s\}$ we are once again
in the conjugate setting. A graphical model for the weak-limit approximation to
the HDP-HSMM including the auxiliary variables is shown in
Figure~\ref{fig:weaklimitfull}.

For a more detailed derivation as well as further numerical experiments, see
\cite{johnson2012truncated}.

\subsection{A Direct Assignment Sampler for the HDP-HSMM}
Though all the experiments in this paper are performed with the weak-limit
sampler, we provide a direct assignment (DA) sampler as well for theoretical
completeness and because it may be useful in cases where there is
insufficient data to inform some latent parameters so that marginalization is
necessary for mixing or estimating marginal likelihoods (such as in some topic
models). As mentioned previously, in the direct assignment sampler for the HDP-HMM the
infinite transition matrix $\pi$ is analytically marginalized out along with
the observation parameters (if conjugate priors are used). The sampler
represents explicit instantiations of the state sequence $(x_t)$ and the
``used'' prefix of the infinite vector $\beta$: $\beta_{1:K}$ where $K=\#\{x_t
    : t =1,\ldots,T\}$. There are also auxiliary variables $m$ used to resample
$\beta$, but for simplicity we do not discuss them here; see
\citep{originalhdp} for details.
% Both of our HDP-HSMM direct assignment
% samplers marginalize the infinite transition matrix and observation parameters
% as well as the duration parameters.

Our DA sampler additionally represents the auxiliary variables necessary to
recover HDP conjugacy (as introduced in the previous section). Note that the
requirement for, and correctness of, the auxiliary variables described in the
finite setting in Section~\ref{sec:finite_conjugacy} immediately extends to the
infinite setting as a consequence of the Dirichlet Process's definition in
terms of the finite Dirichlet distribution and the Kolmogorov extension theorem
\citep[Ch.
4]{ccinlar2010probability}; for a detailed discussion, see
\citep{orbanz2010construction}. The connection to the finite case can also be
seen in the sampling steps of the direct assignment sampler for the HDP-HMM,
in which the global weights $beta$ over $K$ instantiated components are
resampled according to $(\beta_{1:K}, \beta_{\text{rest}}) | \alpha \sim
\text{Dir}(\alpha + n_1, \ldots, \alpha + n_K, \alpha)$ where $n_i$ is the
number of transitions into state $i$ and $\text{Dir}$ is the finite Dirichlet
distribution.

\subsubsection{Resampling $(x_t)$}
As described in \citep{emilythesis}, the basic HDP-HMM DA sampling step for
each element $x_t$ of the label sequence is to sample a new label $k$ with
probability proportional (over $k$) to
\begin{align}
    \label{eqn:hdp-hmm-da-score}
    p(x_t = k | (x_{\setminus t}), \beta) \propto \underbrace{\frac{\alpha
            \beta_{k} + n_{x_{t-1},k}}{\alpha  +
            n_{x_{t-1},\cdot}}}_{\text{left-transition}} \cdot
    \underbrace{\frac{\alpha \beta_{x_{t+1}} + n_{k,x_{t+1}} +
            \indicator[x_{t-1} = k = x_{t+1}]}{\alpha + n_{k,\cdot} +
            \indicator[x_{t-1} = k]}}_{\text{right-transition}} \cdot
    \underbrace{f_{\text{obs}}(y_t|x_t=k)}_{\text{observation}}
\end{align}
for $k=1,\ldots,K+1$ where $K=\#\{x_t : t =1,\ldots,T\}$ and where $\indicator$
is an indicator function taking value $1$ if its argument
condition is true and $0$ otherwise.\footnote{The indicator variables are
    present because the two transition probabilities are not independent but
    rather exchangeable.}
The variables $n_{ij}$ are transition counts
in the portion of the state sequence we are conditioning on, i.e.~$n_{ij} =
\#\{x_\tau=i,x_{\tau+1}=j : \tau \in \{1,\ldots,T-1\} \setminus \{t-1,t\}\}$.
The function $f_{\text{obs}}$ is a predictive likelihood:
\begin{align}
    f_{\text{obs}}(&y_t|k) := p(y_t | x_t = k, \{ y_\tau : x_\tau=k \}, H )
    \\ &= \int_{\theta_k} \underbrace{p(y_t | x_t=k,
        \theta_k)}_{\text{likelihood}} \underbrace{\prod_{\tau : x_\tau = k}
        p(y_\tau | x_\tau=k, \theta_k)}_{\text{likelihood of data with same
            label}} \underbrace{p(\theta_k | H)}_{\text{observation parameter
            prior}} \; \dif \theta_k
\end{align}
 We can derive this step by writing the complete joint probability $p((x_t),
 (y_t) | \beta, H)$ leveraging exchangeability; this joint probability value
 is proportional to the desired posterior probability $p(x_t | (x_{\setminus
     t}), (y_t), \beta, H)$. When we consider each possible assignment $x_t=k$,
 we can cancel all the terms that are invariant over $k$, namely all the
 transition probabilities other than those to and from $x_t$ and all data
 likelihoods other than that for $y_t$.  However, this cancellation process
 relies on the fact that for the HDP-HMM there is no distinction between
 self-transitions and new transitions: the term for each $t$ in the complete
 posterior simply involves transition scores no matter the labels of $x_{t+1}$
 and $x_{t-1}$. In the HDP-HSMM case, we must consider segments and their
 durations separately from transitions.

\begin{figure}[tbp]
    \centering \includegraphics[scale=0.4]{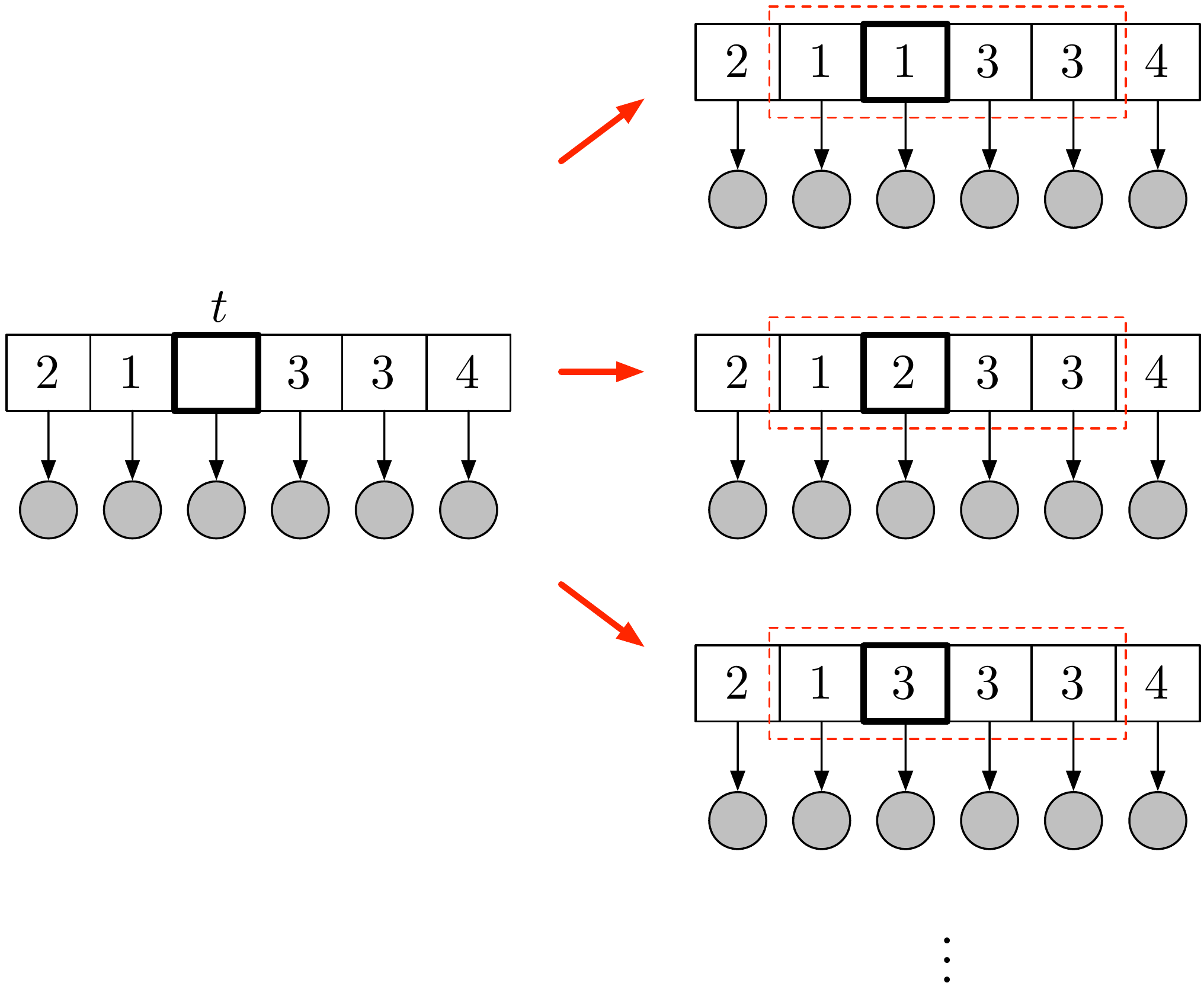}
    \caption{Illustration of the Gibbs step to resample $x_t$ for the
        DA sampler for the HDP-HSMM\@. The red boxes indicate the
        elements of the label sequence that contribute to the score computation
        for $k=1,2,3$ which produce two, three, and two segment terms,
        respectively. The label sequence element being resample is emphasized
        in bold.}
    \label{fig:hdp-hsmm-da-labels}
\end{figure}

To derive an expression for resampling $x_t$ in the case of the HDP-HSMM, we
can similarly consider writing out an expression for the joint probability
$p((x_t), (y_t) | \beta, H, G)$, but we notice that as we vary our
assignment of $x_t$ over $k$, the terms in the expression must change: if
$x_{t-1} = \bar{k}$ or $x_{t+1}=\bar{k}$, the probability expression includes a
segment term for entire contiguous run of label $\bar{k}$. Hence, since we can
only cancel terms that are invariant over $k$, our score expression must
include terms for the adjacent segments into which $x_t$ may merge. See
Figure~\ref{fig:hdp-hsmm-da-labels} for an illustration.

The final expression for the probability of sampling the new value of $x_t$ to be
$k$ then consists of between 1 and 3 segment score terms, depending on merges
with adjacent segments, each of which has the form
\begin{align}
    \label{hdp-hsmm-da-score}
    p(x_t = k | (x_{\setminus t}), \beta, H, G) \propto
    \underbrace{\frac   {\alpha \beta_{k} + n_{x_{\text{prev}},k}}
                        {\alpha (1-\beta_{x_{\text{prev}}}) + n_{x_{\text{prev}},\cdot}}
                    }_{\text{left-transition}}
                    \cdot
    \underbrace{\frac   {\alpha \beta_{x_{\text{next}}} + n_{k,x_{\text{next}}}}
                        {\alpha(1-\beta_{k}) + n_{k,\cdot}}
                    }_{\text{right-transition}} \\
                    \cdot
    \underbrace{f_{\text{dur}}(t^2-t^1+1)
                    }_{\text{duration}}
                    \cdot
    \underbrace{f_{\text{obs}}(y_{t^1:t^2} | k)
                    }_{\text{observation}}
\end{align}
where we have used $t^1$ and $t^2$ to denote the first and last indices of the
segment, respectively.  Transition scores at the start and end of the chain are
not included.

The function $f_{\text{dur}}( d | k)$ is the corresponding duration predictive
likelihood evaluated on a duration $d$, which depends on the durations of other
segments with label $k$ and any duration hyperparameters.  The function
$f_\text{obs}$ now represents a \emph{block} or \emph{joint} predictive
likelihood over all the data in a segment (see
e.g.~\citep{Murphy07conjugatebayesian} for a thorough discussion of the
Gaussian case). Note that the denominators in the transition terms are affected
by the elimination of self-transitions by a rescaling of the ``total mass.''
The resulting chain is ergodic if the duration predictive score
$f_{\text{dur}}$ has a support that includes $\{1,2,\ldots,d_{\text{max}}\}$,
so that segments can be split and merged in any combination.

\subsubsection{Resampling $\beta$ and Auxiliary Variables $\rho$}
\label{sec:hdp-hsmm-da-aux}
To allow conjugate resampling of $\beta$, auxiliary variables must be
introduced to deal with the conjugacy issue raised in
Section~\ref{sec:weaklimit}. In the direct assignment samplers, the auxiliary
variables are not used to resample diagonal entries of the transition matrix
$\pi$, which is marginalized out, but rather to directly resample $\beta$.
In particular, with each segment $s$ we associate an auxiliary count $\rho_s$
which is independent of the data and only serves to preserve conjugacy in the
HDP\@. We periodically re-sample via
\begin{align}
    \pi_{ii} | \alpha, \beta  &\sim \text{Beta}(\alpha \beta_i , \alpha (1-\beta_i) )\\
    \rho_s | \pi_{ii}, z_s &\sim \text{Geo}(1-\pi_{z_s,z_s})
\end{align}
The count $n_{i,i}$, which is used in resampling the auxiliary variables $m$ of
\cite{originalhdp} which in turn are then used to resample $\beta$, is the
total of the auxiliary variables for other segments with the same label, i.e.
$n_{i,i} = \sum_{\bar{s} \neq s, z_{\bar{s}}=z_s} \rho_{\bar{s}}$. This formula
can be interpreted as simply sampling the number of self-transitions we may
have seen at segment $s$ given $\beta$ and the counts of self- and non-self
transitions in the super-state sequence. Note $\pi_{ii}$ is independent of the
data given $(z_s)$; as before, this auxiliary variable procedure is
a convenient way to integrate out numerically the diagonal entries of the
transition matrix.

By using the total auxiliary as the statistics for $n_{i,i}$, we can resample
$\beta | (x_t), \alpha, \gamma$ according to the procedure for the HDP-HMM as
described in \citep{originalhdp}.

\subsection{Exploiting Changepoint Side-Information}
\label{sec:exploiting-changepoints}
In many circumstances, we may not need to consider all time indices as possible
changepoints at which the super-state may switch; it may be easy to rule out
many non-changepoints from consideration. For example, in the power
disaggregation application in Section~\ref{sec:experiments}, we can run
inexpensive changepoint detection on the observations to get a list of
\emph{possible} changepoints, ruling out many obvious non-changepoints. The
possible changepoints divide the label sequence into state \emph{blocks}, where
within each block the label sequence must be constant, though sequential blocks
may have the same label. By only allowing super-state switching to occur at
these detected changepoints, we can greatly reduce the computation of all the
samplers considered.

In the case of the weak-limit sampler, the complexity of the bottleneck
message-passing step is reduced to a function of the number of possible
changepoints (instead of total sequence length): the asymptotic complexity
becomes $\mathcal{O}(T_\text{change}^2N + N^2 T_\text{change})$
, where $T_\text{change}$, the number of possible changepoints, may
be dramatically smaller than the sequence length $T$. We simply modify the
backwards message-passing procedure to sum only over the possible durations:
\begin{align}
    B_t^*(i) &:= p(y_{t+1:T} | x_{t+1} = i, F_t = 1)\\
    &= \sum_{d \in \mathbb{D}}^{} B_{t + d}(i) \underbrace{\tilde{p}(D_{t+1}=d|x_{t+1} = i)}_{\text{duration prior term}} 
    \cdot \underbrace{p(y_{t+1:t+d} | x_{t+1}=i, D_{t+1}=d)}_{\text{likelihood term}} \\
    &\quad + \underbrace{\tilde{p}(D_{t+1} > T-t | x_{t+1}=i) p(y_{t+1:T} | x_{t+1}=i, D_{t+1}>T-t)}_{\text{censoring term}}
\end{align}
where $\tilde{p}$ represents the duration distribution restricted to the set of
possible durations $\mathbb{D} \subset \mathbb{N}^+$ and re-normalized. We
similarly modify the forward-sampling procedure to only consider possible
durations.
It is also clear how to adapt the DA sampler: instead of re-sampling each
element of the label sequence $(x_t)$ we simply consider the block label
sequence, resampling each block's label (allowing merging with adjacent
blocks).

\section{Experiments}
\label{sec:experiments}
In this section, we evaluate the proposed HDP-HSMM sampling algorithms on both
synthetic and real data. First, we compare the HDP-HSMM direct assignment
sampler to the weak limit sampler as well as the Sticky HDP-HMM direct
assignment sampler, showing that the HDP-HSMM direct assignment sampler has
similar performance to that for the Sticky HDP-HMM and that the weak limit
sampler is much faster. Next, we evaluate the HDP-HSMM weak limit sampler on
synthetic data generated from finite HSMMs and HMMs. We show that the HDP-HSMM
applied to HSMM data can efficiently learn the correct model, including the
correct number of states and state labels, while the HDP-HMM is unable to
capture non-geometric duration statistics. We also apply the HDP-HSMM to data
generated by an HMM and demonstrate that, when equipped with a duration
distribution class that includes geometric durations, the HDP-HSMM can also
efficiently learn an HMM model when appropriate with little loss in efficiency.
Next, we use the HDP-HSMM in a factorial \citep{ghahramani1997factorial}
structure for the purpose of disaggregating a whole-home power signal into the
power draws of individual devices. We show that encoding simple duration prior
information when modeling individual devices can greatly improve performance,
and further that a Bayesian treatment of the parameters is advantageous. We
also demonstrate how changepoint side-information can be leveraged to
significantly speed up computation. The Python code used to perform these
experiments as well as Matlab code is available online at
\url{http://github.com/mattjj/pyhsmm}.

\subsection{Synthetic Data}
\begin{figure}
    \centering
    \subfigure[]{\includegraphics[width=2.5in]{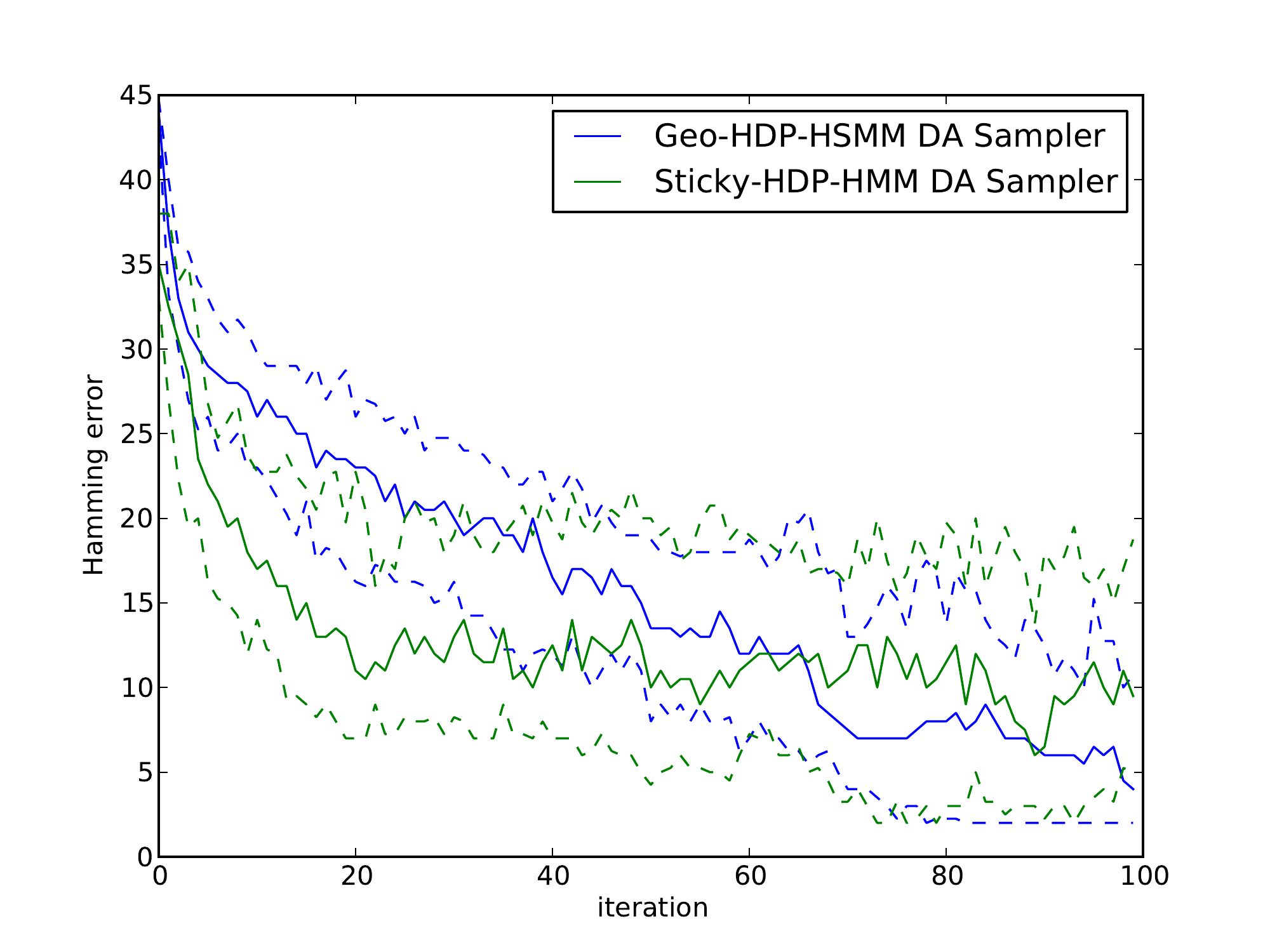}
        \label{fig:hsmm_vs_stickyhmm}}
    \subfigure[]{\includegraphics[width=2.5in]{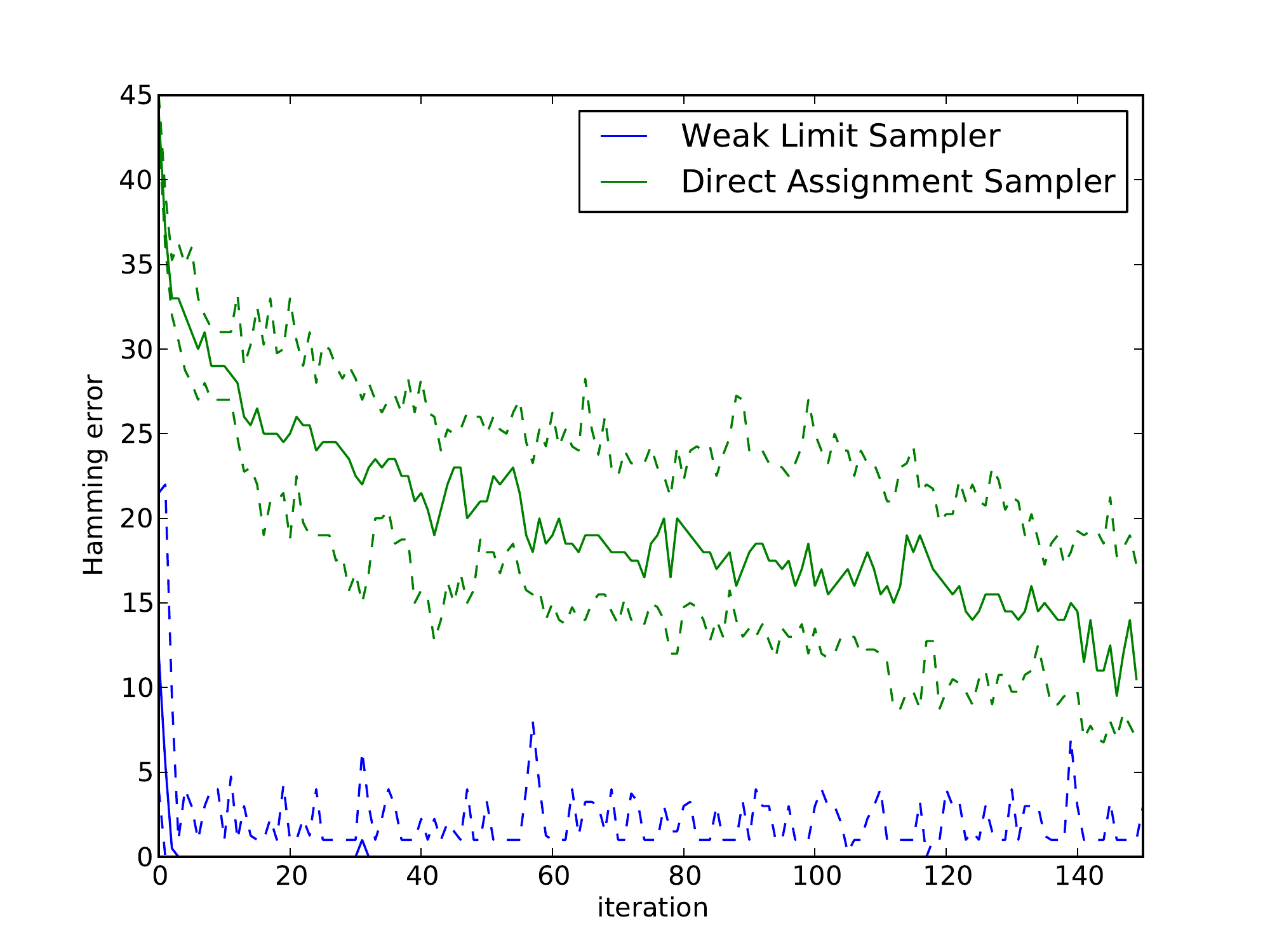}
        \label{fig:wl_is_faster_hamming}}
    \caption{\ref{fig:hsmm_vs_stickyhmm} compares the Geometric-HDP-HSMM direct
        assignment sampler with that of the Sticky HDP-HMM, both applied to HMM
        data. The sticky parameter $\kappa$ was chosen to maximize
        mixing.~\ref{fig:wl_is_faster_hamming}
         compares the HDP-HSMM direct
        assignment sampler with the weak limit sampler. In all plots, solid lines are the median
        error at each time over 25 independent chains; dashed lines are 25th
        and 75th percentile errors.}
    \label{fig:hsmm_vs_sticky_wl_is_faster}
\end{figure}

Figure~\ref{fig:hsmm_vs_sticky_wl_is_faster} compares the HDP-HSMM direct
assignment sampler to that of the Sticky HDP-HMM as well as the HDP-HSMM weak
limit sampler. Figure~\ref{fig:hsmm_vs_stickyhmm} shows that the direct
assignment sampler for a Geometric-HDP-HSMM performs similarly to the Sticky
HDP-HSMM direct assignment sampler when applied to data generated by an HMM
with scalar Gaussian emissions. Figures~\ref{fig:wl_is_faster_hamming}
shows that the weak limit sampler mixes
much more quickly than the direct assignment sampler. Each iteration of the
weak limit sampler is also much faster to execute (approximately 50x faster in
our implementations in Python). Due to its much greater efficiency, we focus on the weak
limit sampler for the rest of this section; we believe it is a superior
inference algorithm whenever an adequately large approximation parameter $L$
can be chosen a priori.

\begin{figure}[tp]
    \centering
    \subfigure[1][HDP-HMM]{\includegraphics[height=1.75in]{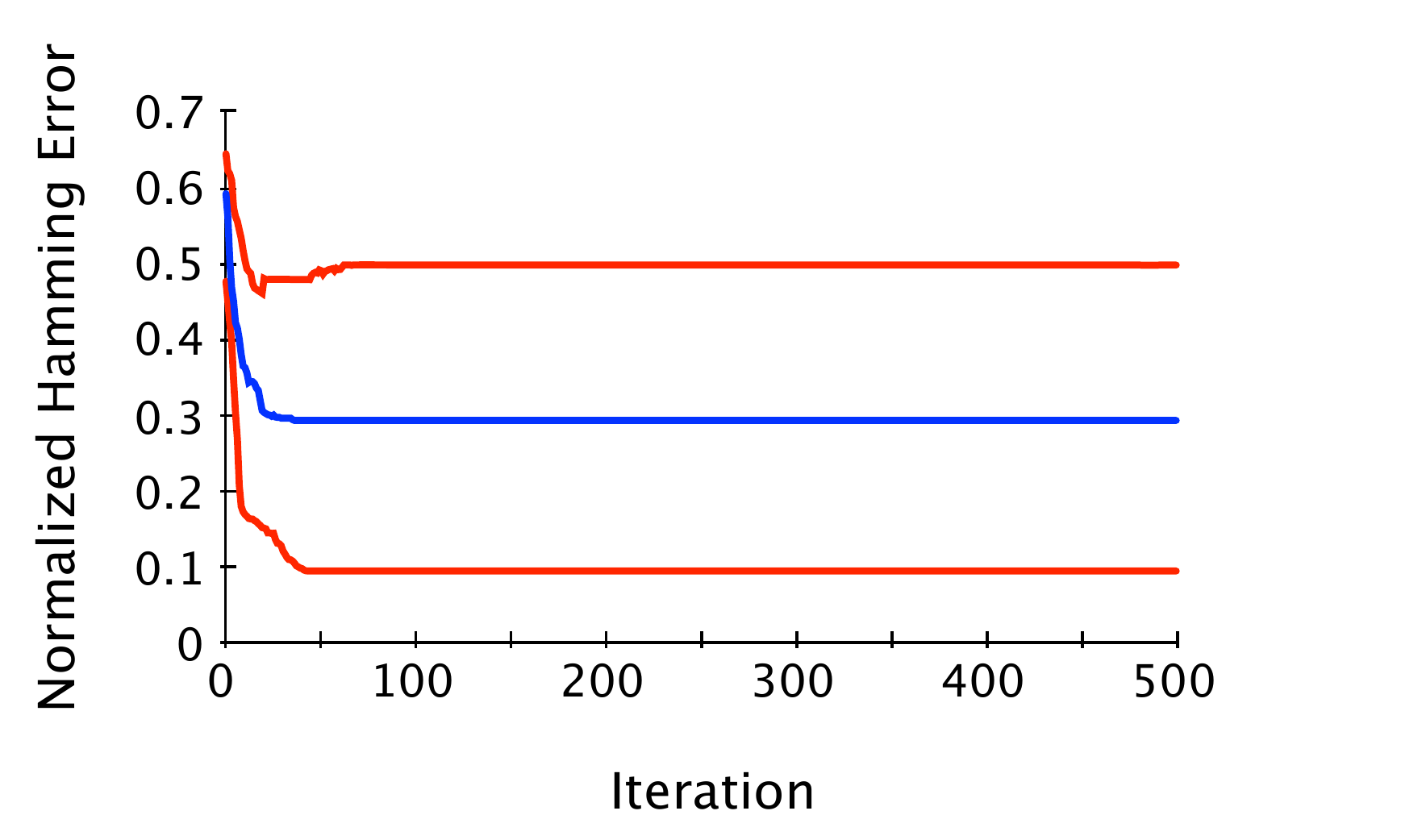}}
    \subfigure[2][HDP-HSMM]{\includegraphics[height=1.75in]{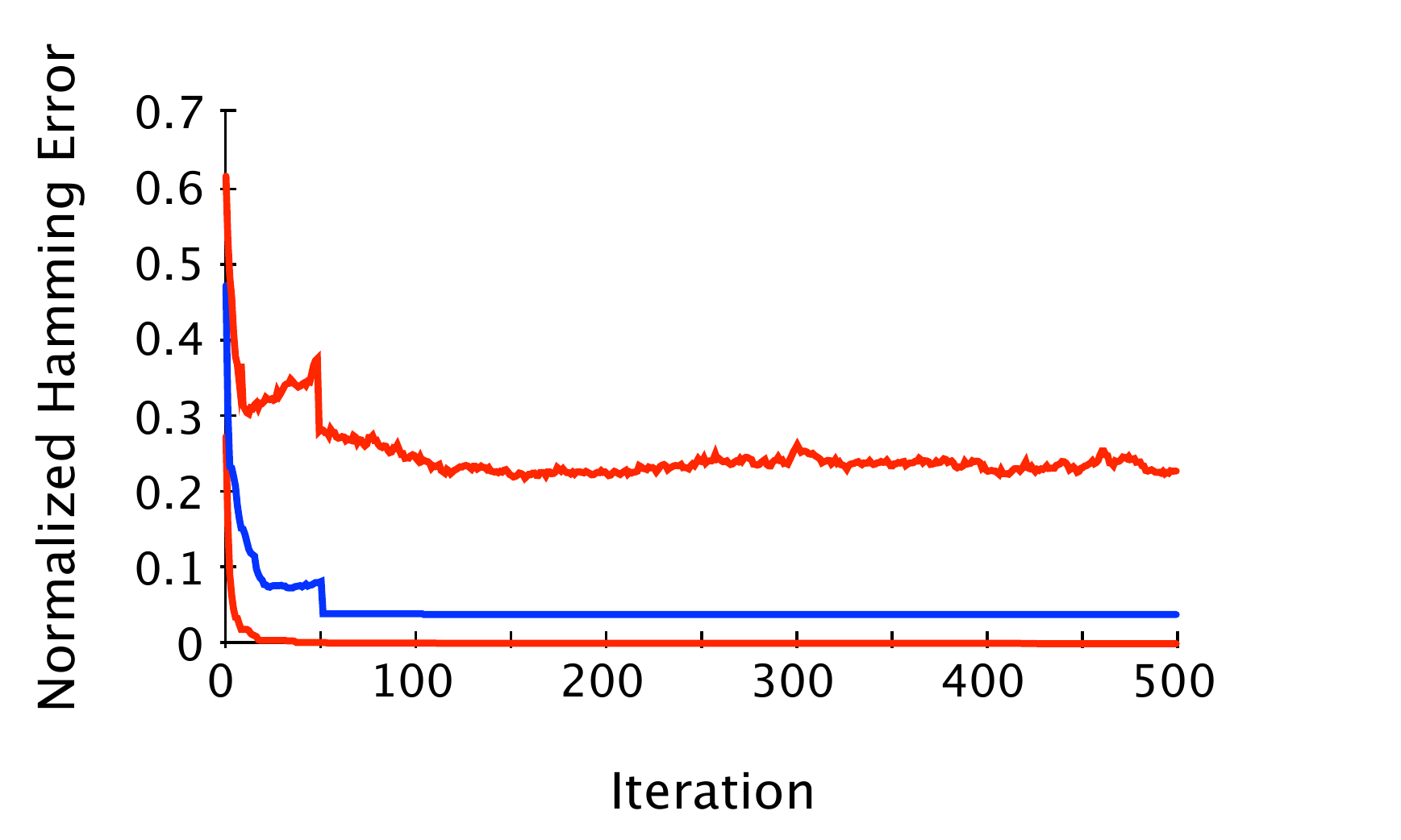}}
    \caption{State-sequence Hamming error of the HDP-HMM and Poisson-HDP-HSMM
        applied to data from a Poisson-HSMM\@. In each plot, the blue line
        indicates the error of the chain with the median error across 25
        independent Gibbs chains, while the red lines indicate the chains with
        the 10th and 90th percentile errors at each iteration. The jumps in the
        plot correspond to a change in the ranking of the 25 chains.}
    \label{fig:on_synthetic_hsmm}
\end{figure}

\begin{figure}[tp]
    \centering
    \subfigure[1][HDP-HSMM]{\includegraphics[height=1.75in]{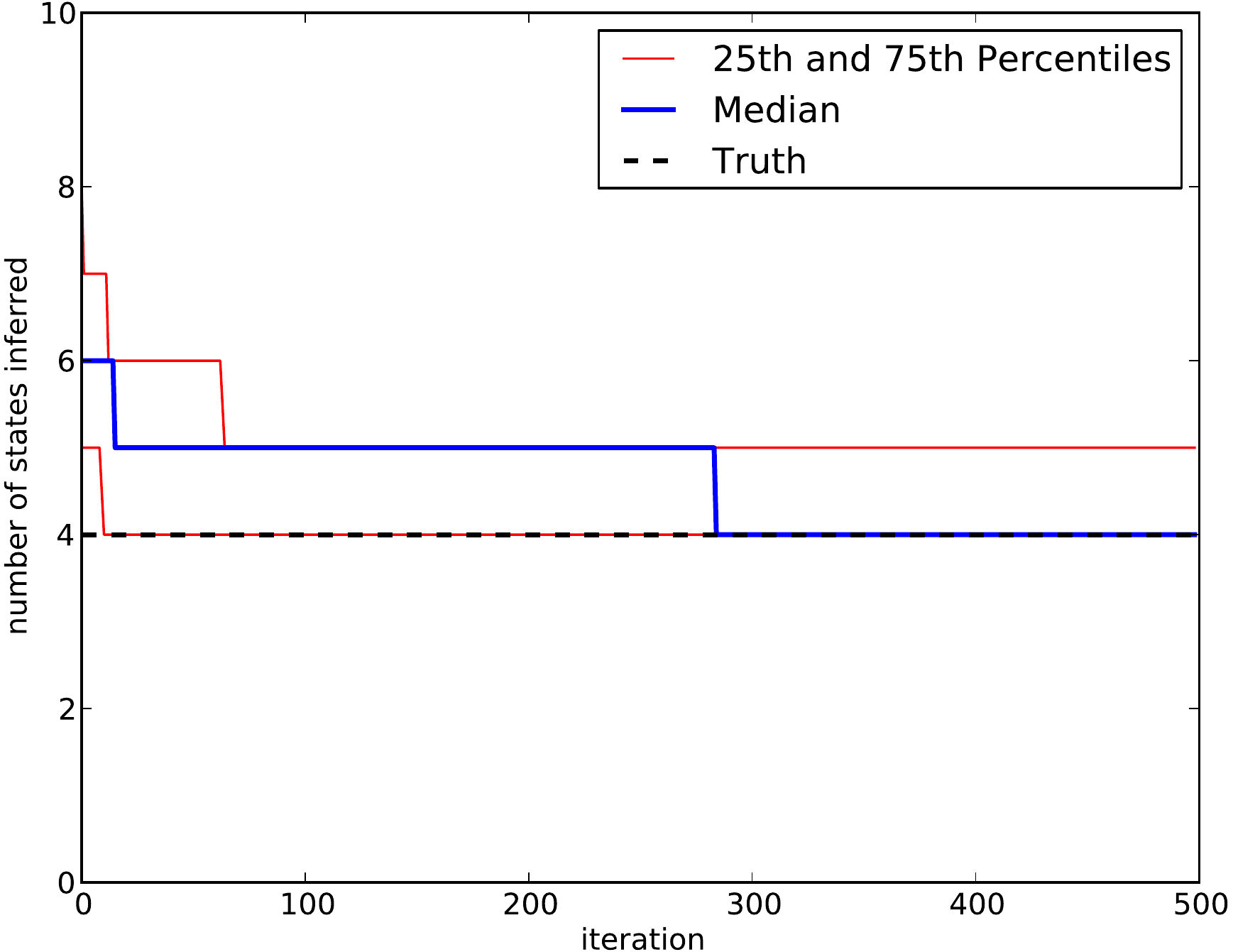}}
    \subfigure[2][HDP-HMM]{\includegraphics[height=1.75in]{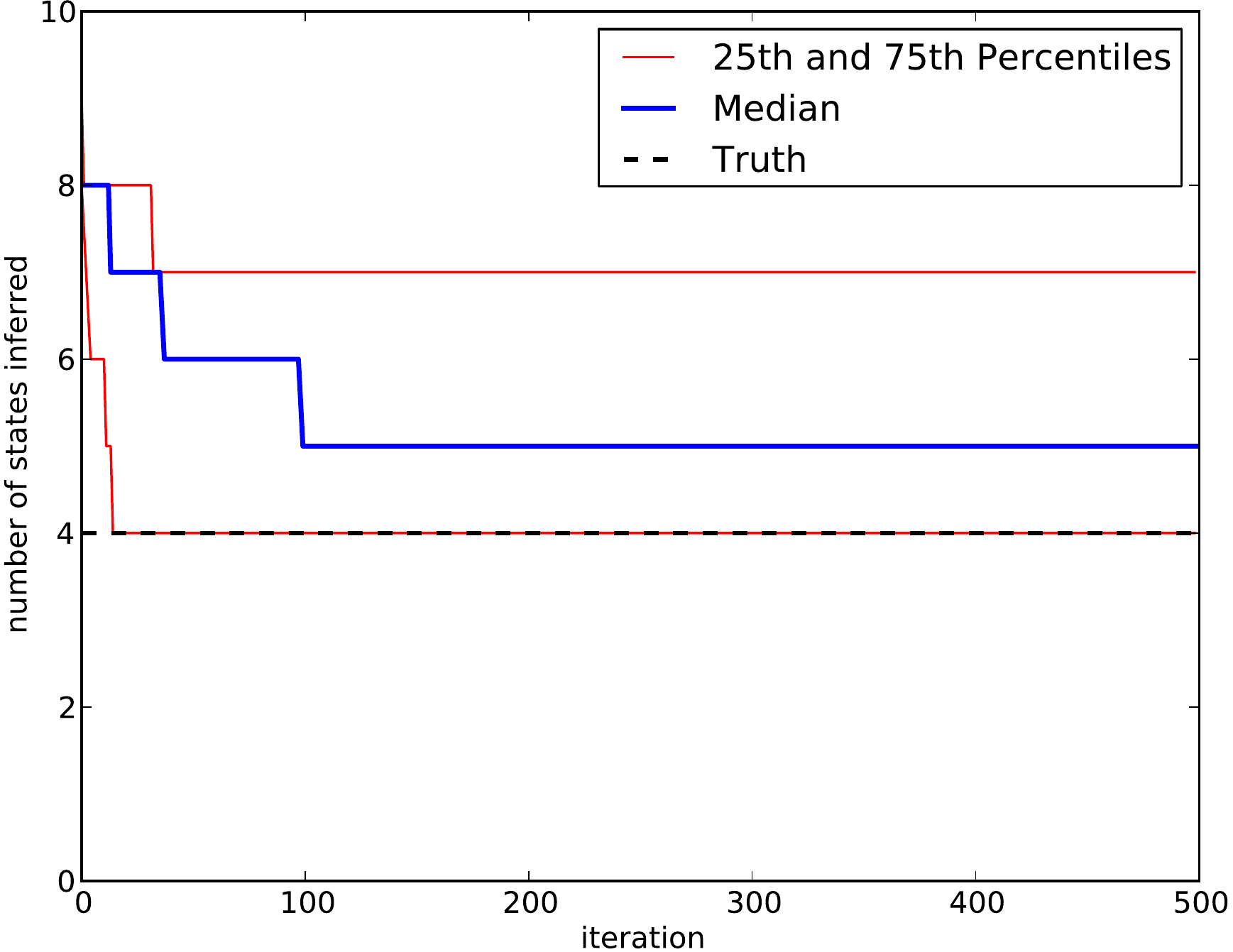}}
    \caption{Number of states inferred by the HDP-HMM and Poisson-HDP-HSMM
        applied to data from a four-state Poisson-HSMM\@. In each plot, the blue
        line indicates the error of the chain with the median error across 25
        independent Gibbs chains, while the red lines indicate the chains with
        the 10th and 90th percentile errors at each iteration.}
    \label{fig:on_synthetic_hsmm_numstates}
\end{figure}

Figure~\ref{fig:on_synthetic_hsmm} summarizes the results of applying both a
Poisson-HDP-HSMM and an HDP-HMM to data generated from an HSMM with four
states, Poisson durations, and 2-dimensional mixture-of-Gaussian emissions. In
the 25 Gibbs sampling runs for each model, we applied 5 chains to each of 5
generated observation sequences.  The HDP-HMM is unable to capture the
non-Markovian duration statistics and so its state sampling error remains high,
while the HDP-HSMM equipped with Poisson duration distributions is able to
effectively learn the correct temporal model, including duration, transition,
and emission parameters, and thus effectively separate the states and
significantly reduce posterior uncertainty. The HDP-HMM also frequently fails
to identify the true number of states, while the posterior samples for the
HDP-HSMM concentrate on the true number; see
Figure~\ref{fig:on_synthetic_hsmm_numstates}.

\begin{figure}[tp]
    \centering
    \subfigure[1][HDP-HMM]{\includegraphics[height=1.7in]{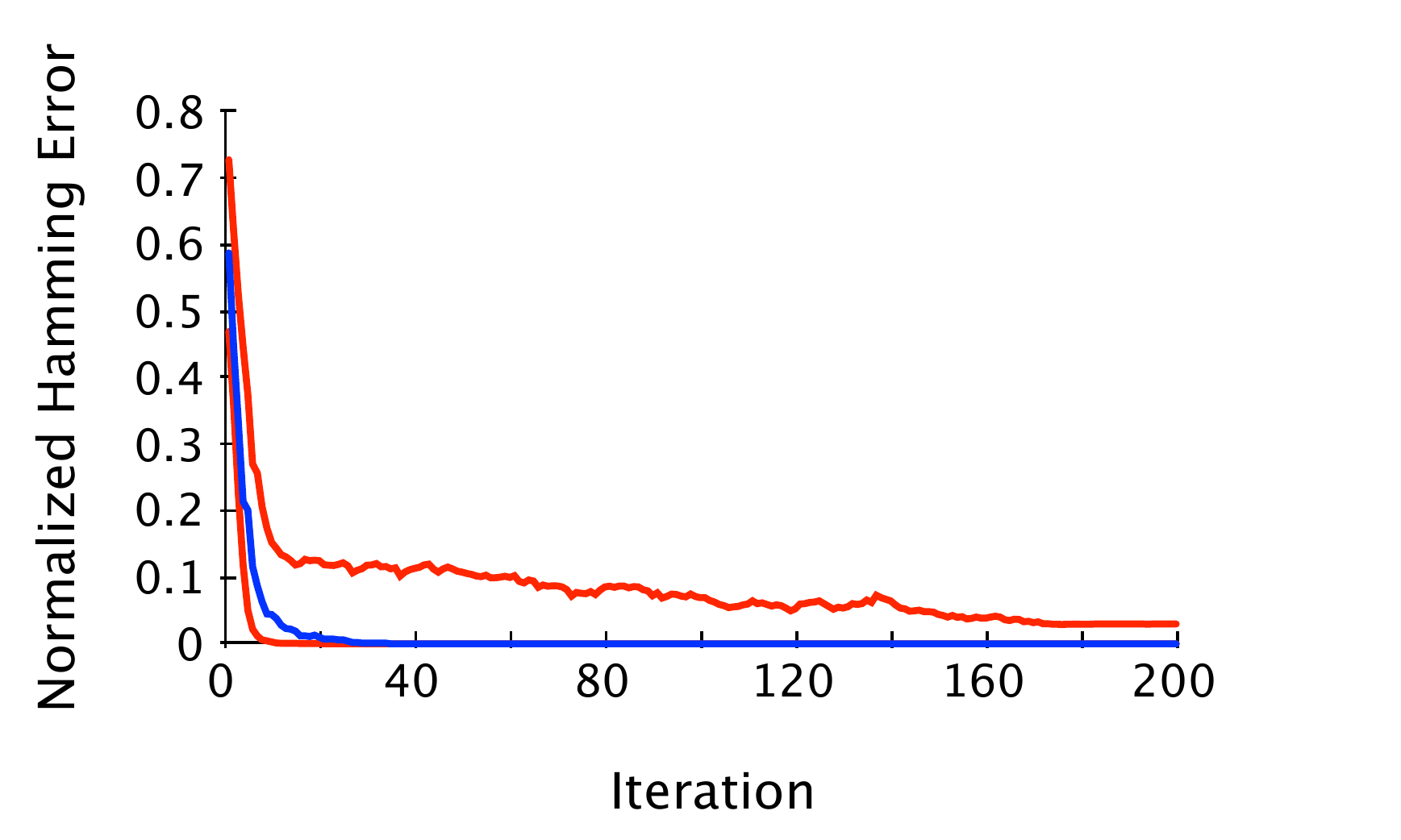}}
    \subfigure[2][HDP-HSMM]{\includegraphics[height=1.7in]{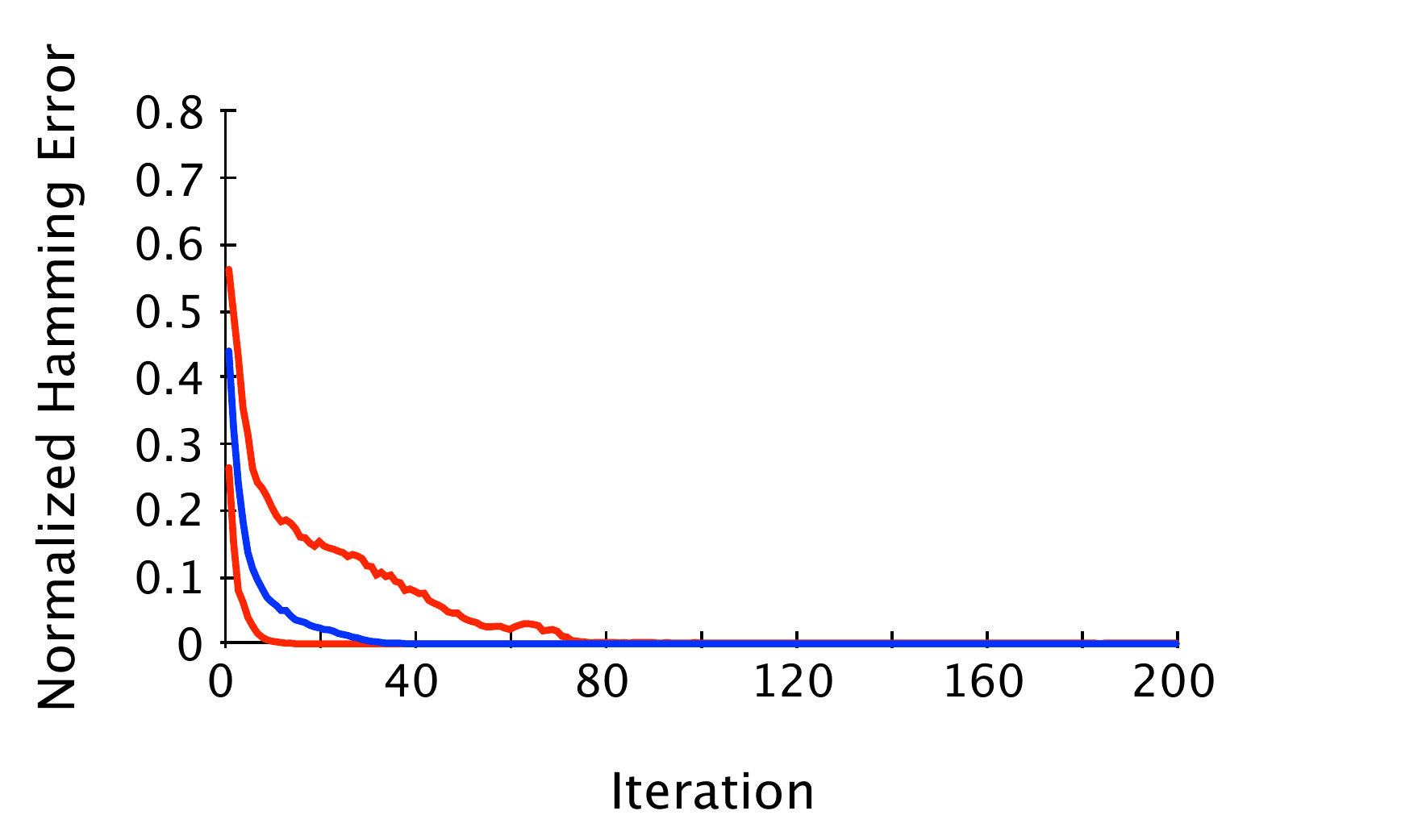}}
    \caption{The HDP-HSMM and HDP-HMM applied to data from an HMM\@. In each
        plot, the blue line indicates the error of the chain with the median
        error across 25 independent Gibbs chains, while the red line indicates
        the chains with the 10th and 90th percentile error at each iteration.}
    \label{fig:on_synthetic_hmm}
\end{figure}

By setting the class of duration distributions to be a superclass of the class
of geometric distributions, we can allow an HDP-HSMM model to learn an HMM from
data when appropriate. One such distribution class is the class of negative
binomial distributions, denoted $\text{NegBin}(r,p)$, the discrete analog of
the Gamma distribution, which covers the class of geometric distributions when
$r=1$. By placing a (non-conjugate) prior over $r$ that includes $r=1$ in its
support, we allow the model to learn geometric durations as well as
significantly non-geometric distributions with modes away from zero.
Figure~\ref{fig:on_synthetic_hmm} shows a negative binomial HDP-HSMM learning
an HMM model from data generated from an HMM with four states. The observation
distribution for each state is a 10-dimensional Gaussian, again with parameters
sampled i.i.d.~from a NIW prior. The prior over $r$ was set to be uniform on
$\{1,2,\ldots,6\}$, and all other priors were chosen to be similarly
non-informative. The sampler chains quickly concentrated at $r=1$ for all state
duration distributions. There is only a slight loss in mixing time for the
HDP-HSMM compared to the HDP-HMM\@. This experiment demonstrates that with the
appropriate choice of duration distribution the HDP-HSMM can effectively learn
an HMM model.

\subsection{Power Disaggregation}
\label{sec:power-disaggregation}
In this section we show an application of the HDP-HSMM factorial structure to
an unsupervised power signal disaggregation problem. The task is to estimate
the power draw from individual devices, such as refrigerators and microwaves,
given an aggregated whole-home power consumption signal. This disaggregation
problem is important for energy efficiency: providing consumers with detailed
power use information at the device level has been shown to improve efficiency
significantly, and by solving the disaggregation problem one can provide that
feedback without instrumenting every individual device with monitoring
equipment. This application demonstrates the utility of including duration
information in priors as well as the significant speedup achieved with
changepoint-based inference.

% \begin{figure}[tp]
%     \centering
%     \includegraphics[width=5in]{figures/disagg_experiments/area_vs_time.pdf}
%     \caption{Power signals recorded for one home in the REDD dataset over a 24
%         hour period. The power disaggregation problem is to estimate the
%         component power signals from the aggregate (sum) signal. Reproduced
%         from \citep{kolter2011redd}; the devices shown do not correspond to
%         those in our experiments.}
%     \label{fig:disagg_overview}
% \end{figure}

The power disaggregation problem has a rich history
\citep{zeifman2011nonintrusive} with many proposed approaches for a variety of
problem specifications. Some recent work \citep{kim2010unsupervised} has
considered applying factorial HSMMs to the disaggregation problem using an EM
algorithm; our work here is distinct in that (1) we do not use training data to
learn device models but instead rely on simple prior information and learn the
model details during inference, (2) our states are not restricted to binary
values and can model multiple different power modes per device, and (3) we use
Gibbs sampling to learn all levels of the model. The work in
\citep{kim2010unsupervised} also explores many other aspects of the problem,
such as additional data features, and builds a very compelling complete
solution to the disaggregation problem, while we focus on the factorial time
series modeling itself.

For our experiments, we used the REDD dataset \citep{kolter2011redd}, which
monitors many homes at high frequency and for extended periods of time. We
chose the top 5 power-drawing devices (refrigerator, lighting, dishwasher,
microwave, furnace) across several houses and identified 18 24-hour segments
across 4 houses for which many (but not always all) of the devices switched on
at least once. We applied a 20-second median filter to the data, and each
sequence is approximately 5000 samples long.

We constructed simple priors that set the rough power draw levels and duration
statistics of the modes for several devices. For example, the power draw from
home lighting changes infrequently and can have many different levels, so an
HDP-HSMM with a bias towards longer negative-binomial durations is appropriate.
On the other hand, a refrigerator's power draw cycle is very regular and
usually exhibits only three modes, so our priors biased the refrigerator
HDP-HSMM to have fewer modes and set the power levels accordingly. For details
on our prior specification, see Appendix~\ref{appendix:power-priors}.  
We did not truncate the duration distributions during inference, and we set the
weak limit approximation parameter $L$ to be twice the number of expected modes
for each device; e.g., for the refrigerator device we set $L=6$ and for
lighting we set $L=20$. We performed sampling inference independently on each
observation sequence.

As a baseline for comparison, we also constructed a factorial sticky HDP-HMM
\citep{emilysticky} with the same observation priors and with duration biases
that induced the same average mode durations as the corresponding HDP-HSMM
priors. We also compare to the factorial HMM performance presented in
\citep{kolter2011redd}, which fit device models using an EM algorithm on
training data.  For the Bayesian models, we performed inference separately on
each aggregate data signal.

\begin{figure}[tp]
    \centering
    \includegraphics[scale=0.5]{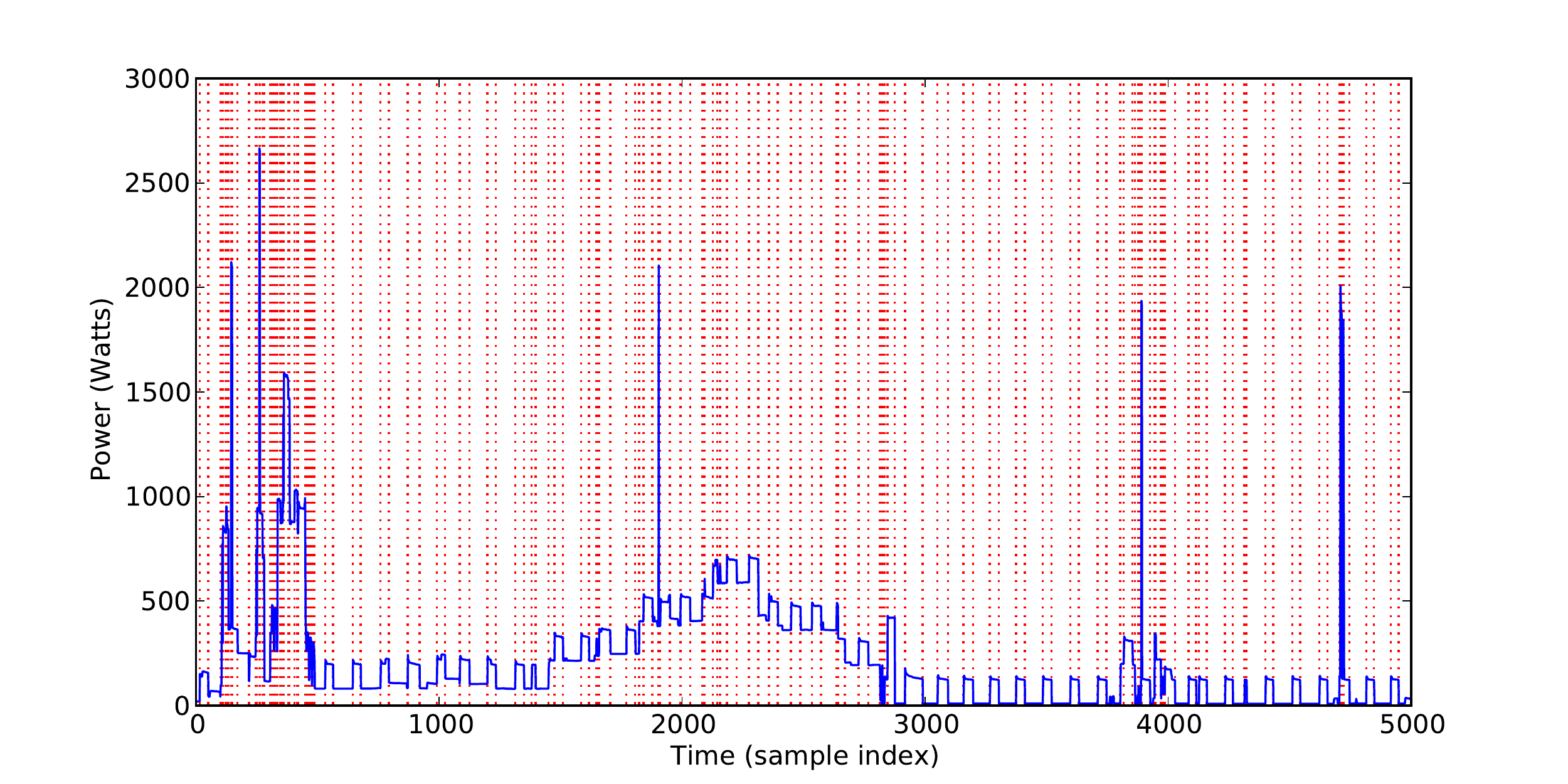}
    \caption{An total power observation sequence from the power disaggregation
        dataset.  Vertical dotted red lines indicate changepoints detected with
        a simple first-differences. By using the changepoint-based algorithms
        described in Section~\ref{sec:exploiting-changepoints} we can greatly
        accelerate inference speed for this application.}
    \label{fig:detected-changepoints}
\end{figure}

The set of possible changepoints is easily identifiable in these data, and a
primary task of the model is to organize the jumps observed in the observations
into an explanation in terms of the individual device models. By simply
computing first differences and thresholding, we are able to reduce the number
of potential changepoints we need to consider from 5000 to 100-200, and hence
we are able to speed up state sequence resampling by orders of magnitude. See
Figure~\ref{fig:detected-changepoints} for an illustration.

To measure performance, we used the error metric of \citep{kolter2011redd}:
\[  \text{Acc.} = 1 - \frac{\sum_{t=1}^T \sum_{i=1}^K \left| \hat{y}_t^{(i)} - y_t^{(i)}\right|}{2 \sum_{t=1}^T \bar{y}_t} \]
where $\bar{y}_t$ refers to the observed total power consumption at time $t$,
$y^{(i)}_t$ is the true power consumed at time $t$ by device $i$, and
$\hat{y}_t^{(i)}$ is the estimated power consumption.  We produced 20 posterior
samples for each model and report the median accuracy of the component emission
means compared to the ground truth provided in REDD\@.
We ran our experiments on standard desktop machines (Intel Core i7-920 CPUs,
released Q4 2008), and a sequence with about 200 detected changepoints would
resample each component chain in 0.1 seconds, including block sampling the
state sequence and resampling all observation, duration, and transition
parameters. We collected samples after every 50 such iterations.

\begin{figure}[tp]
    \centering
    \includegraphics[width=2.5in]{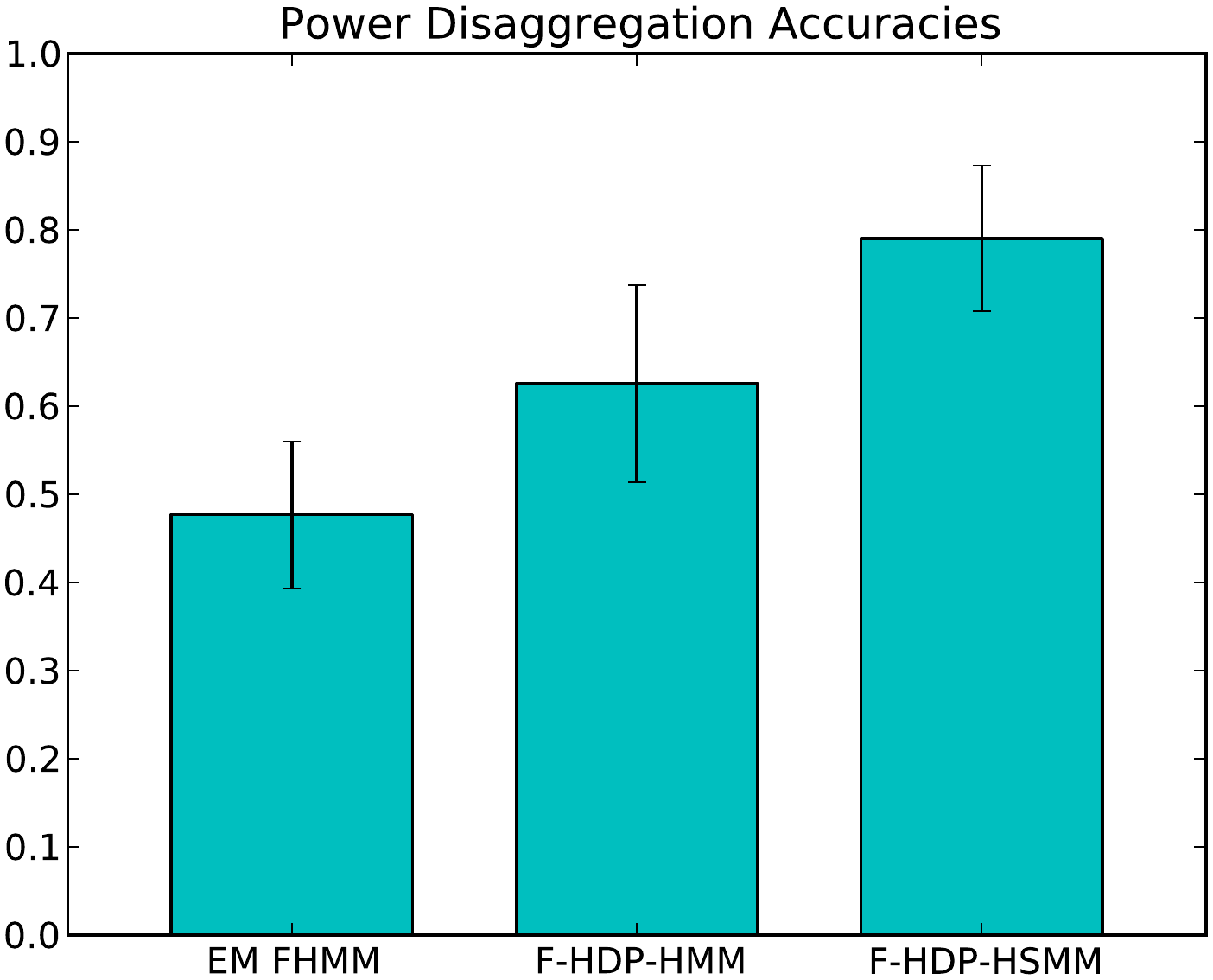}
    \caption{Overall accuracy comparison between the EM-trained FHMM of
        \citep{kolter2011redd}, the factorial sticky HDP-HMM, and the factorial
        HDP-HSMM.}
    \label{fig:disagg_bar}
\end{figure}

\begin{figure}[tp]
    \centering
    \includegraphics[width=2.8in]{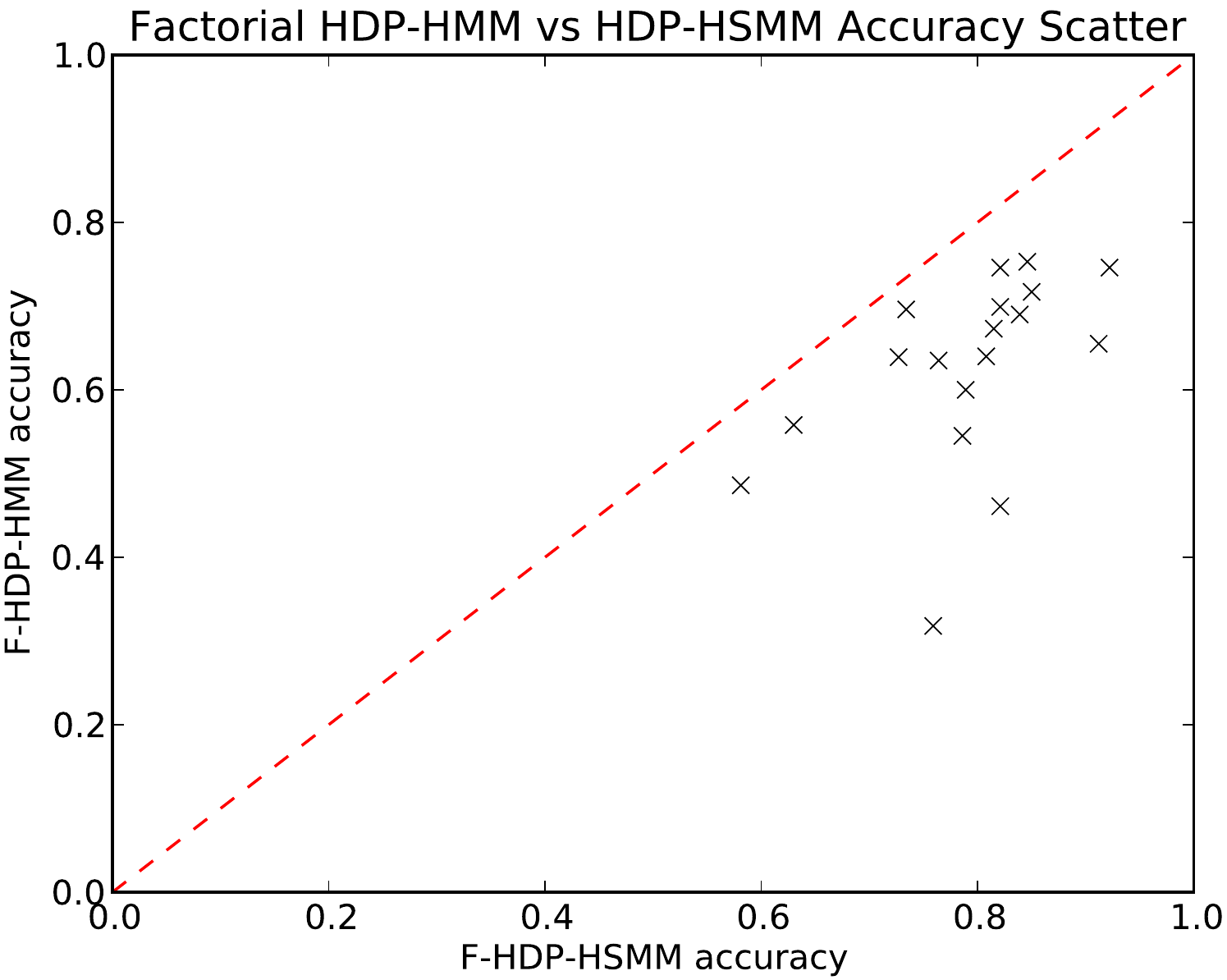}
    \caption{Performance comparison between the HDP-HMM and HDP-HSMM approaches
        broken down by data sequence.}
    \label{fig:disagg_seq_breakdown}
\end{figure}

\begin{figure}[tp]
    \centering
    \subfigure[1][]{\includegraphics[height=1.4in]{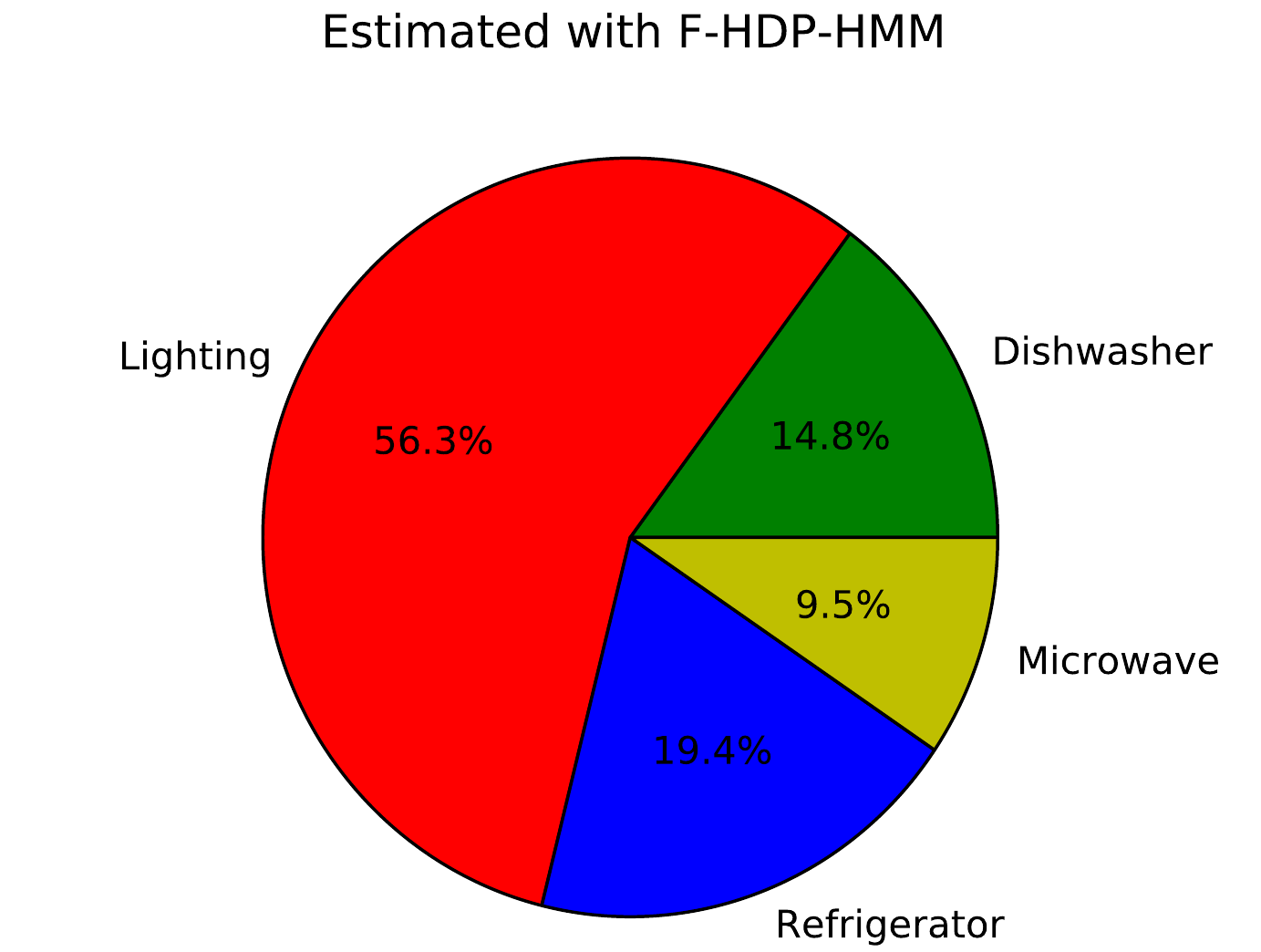}}
    \subfigure[2][]{\includegraphics[height=1.4in]{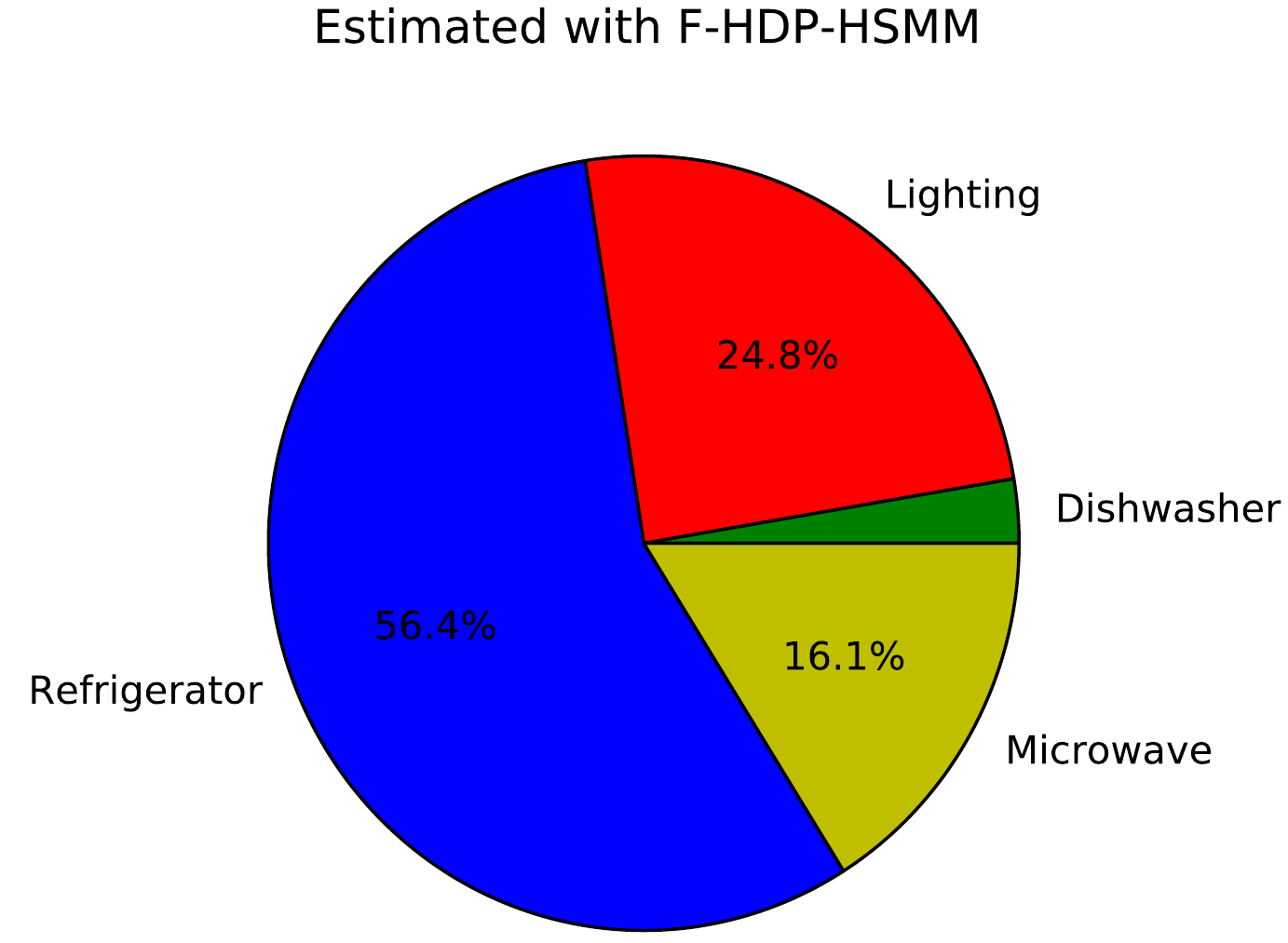}}
    \subfigure[3][]{\includegraphics[height=1.4in]{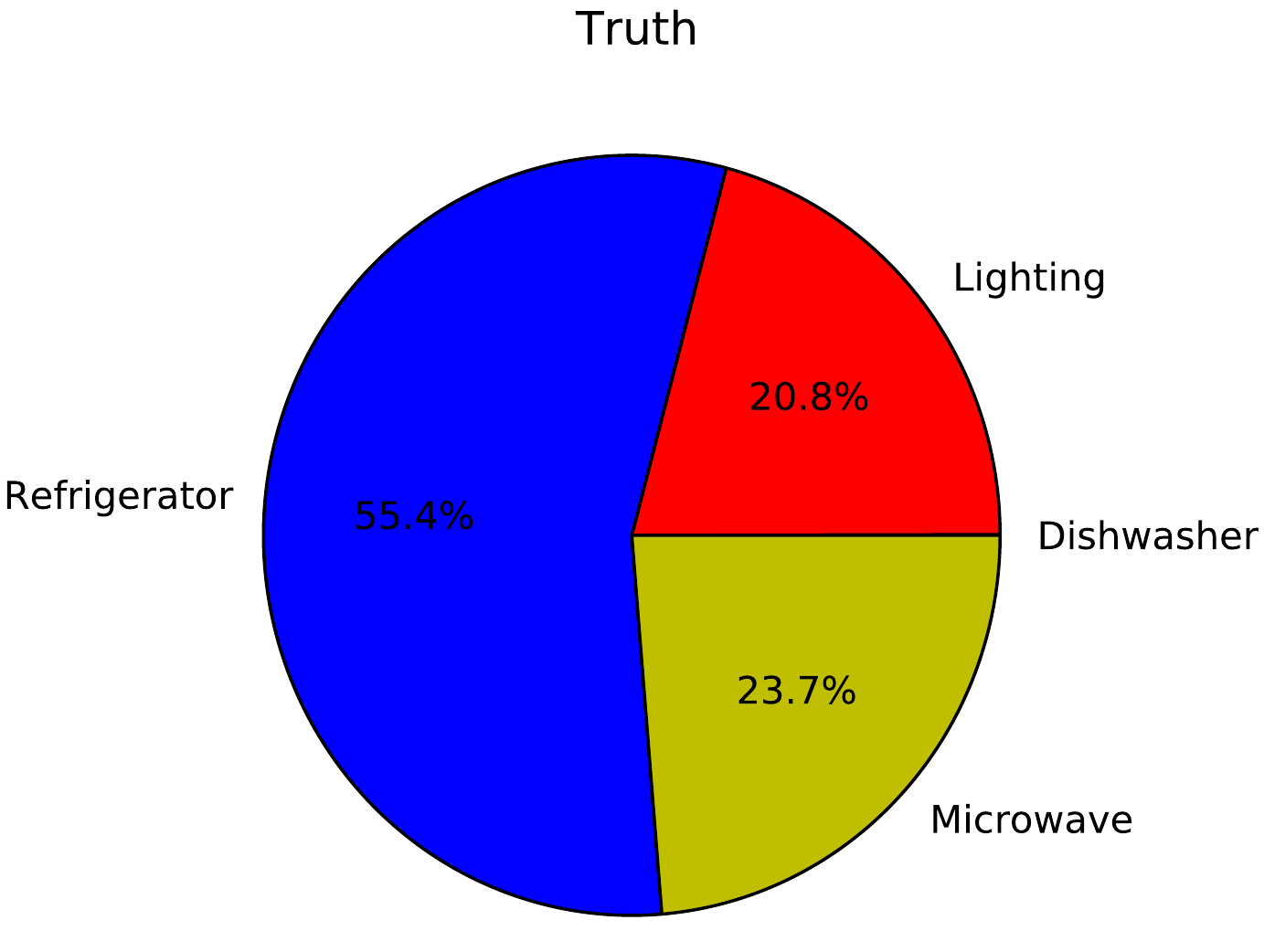}}
    \caption{Estimated total power consumption for a data sequence where the HDP-HSMM significantly outperformed the HDP-HMM due to its modeling of duration regularities.}
    \label{fig:disagg_pies_different}
\end{figure}

\begin{figure}[tp]
    \centering
    \subfigure[1][]{\includegraphics[height=1.4in]{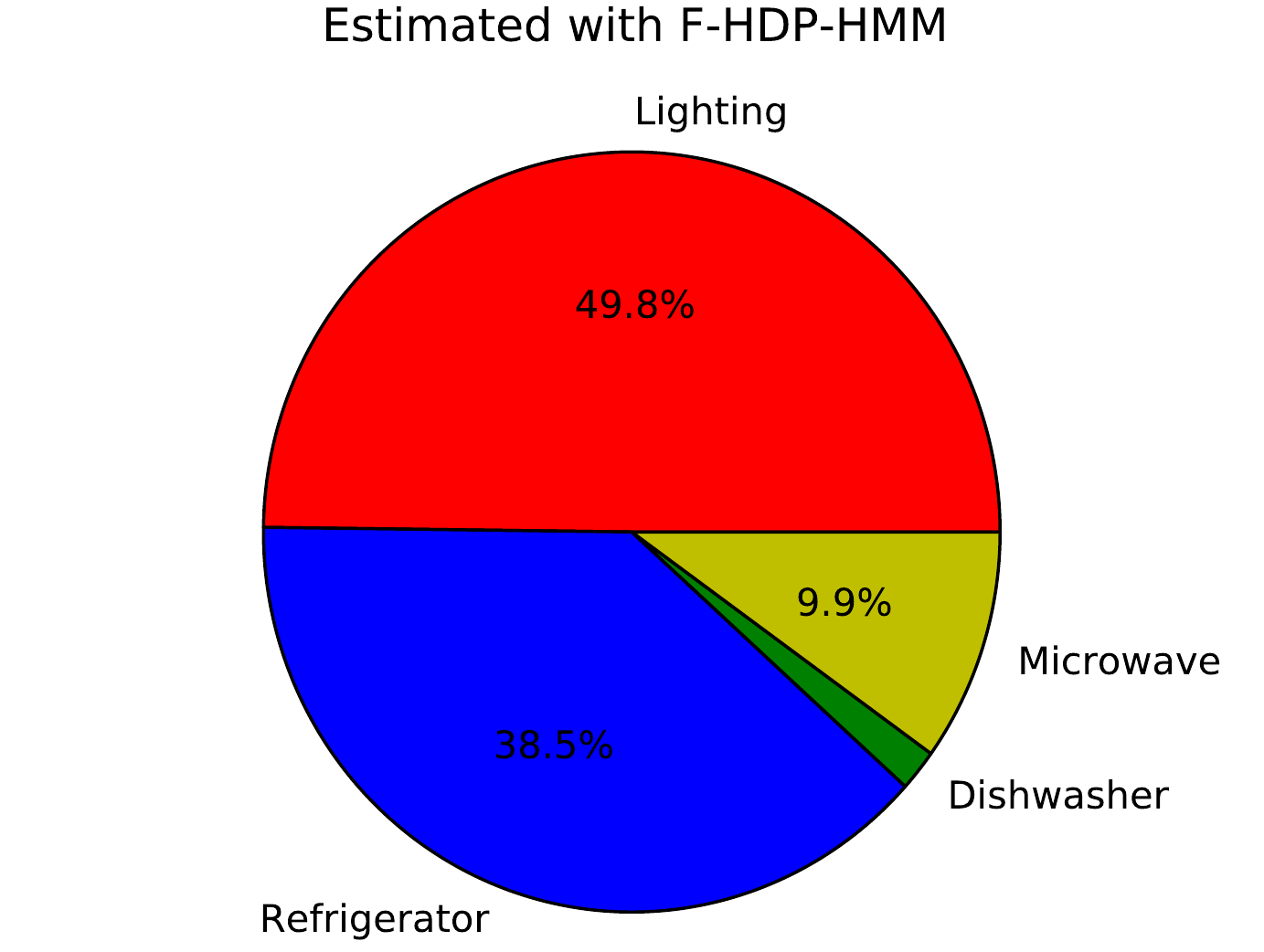}}
    \subfigure[2][]{\includegraphics[height=1.4in]{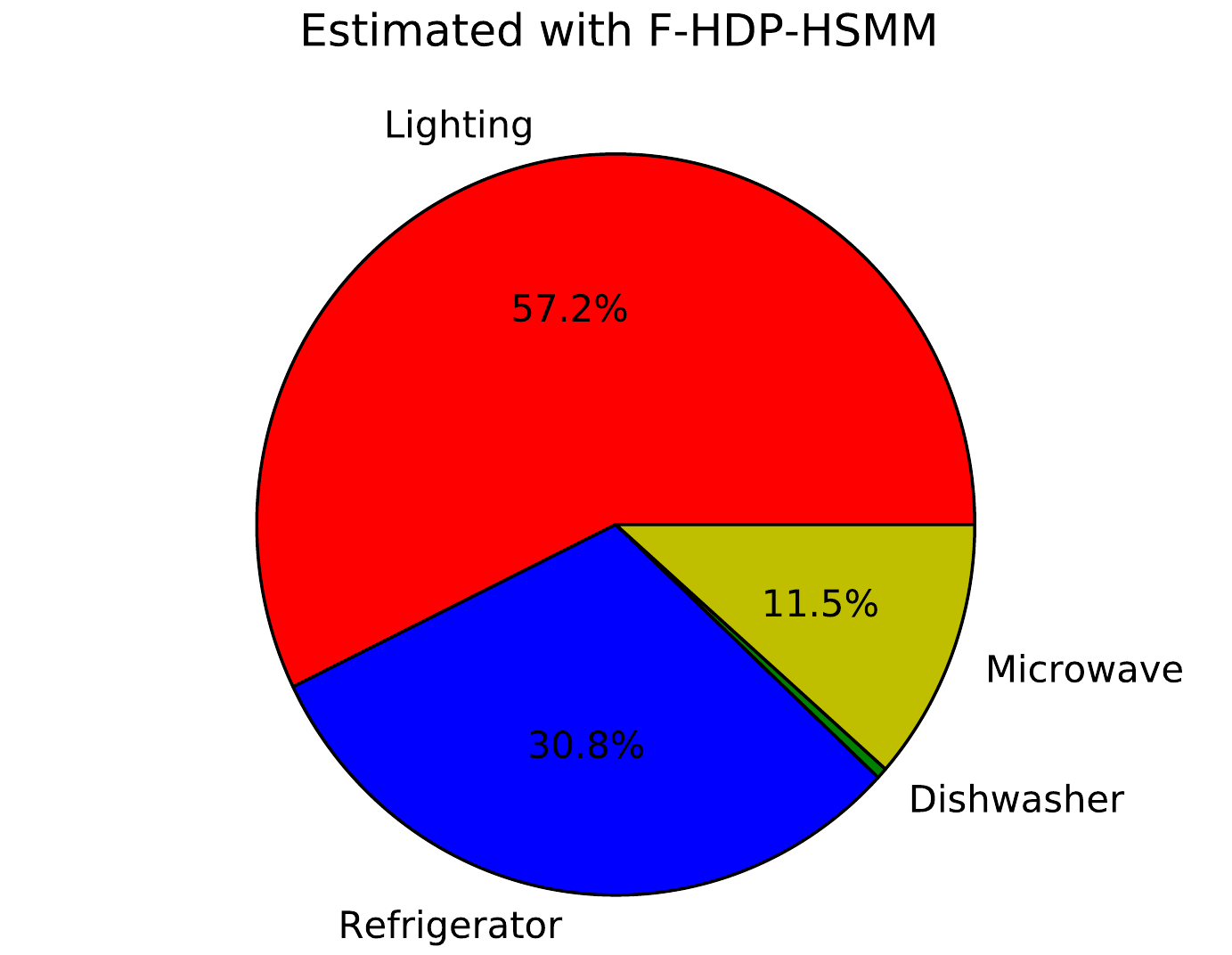}}
    \subfigure[3][]{\includegraphics[height=1.4in]{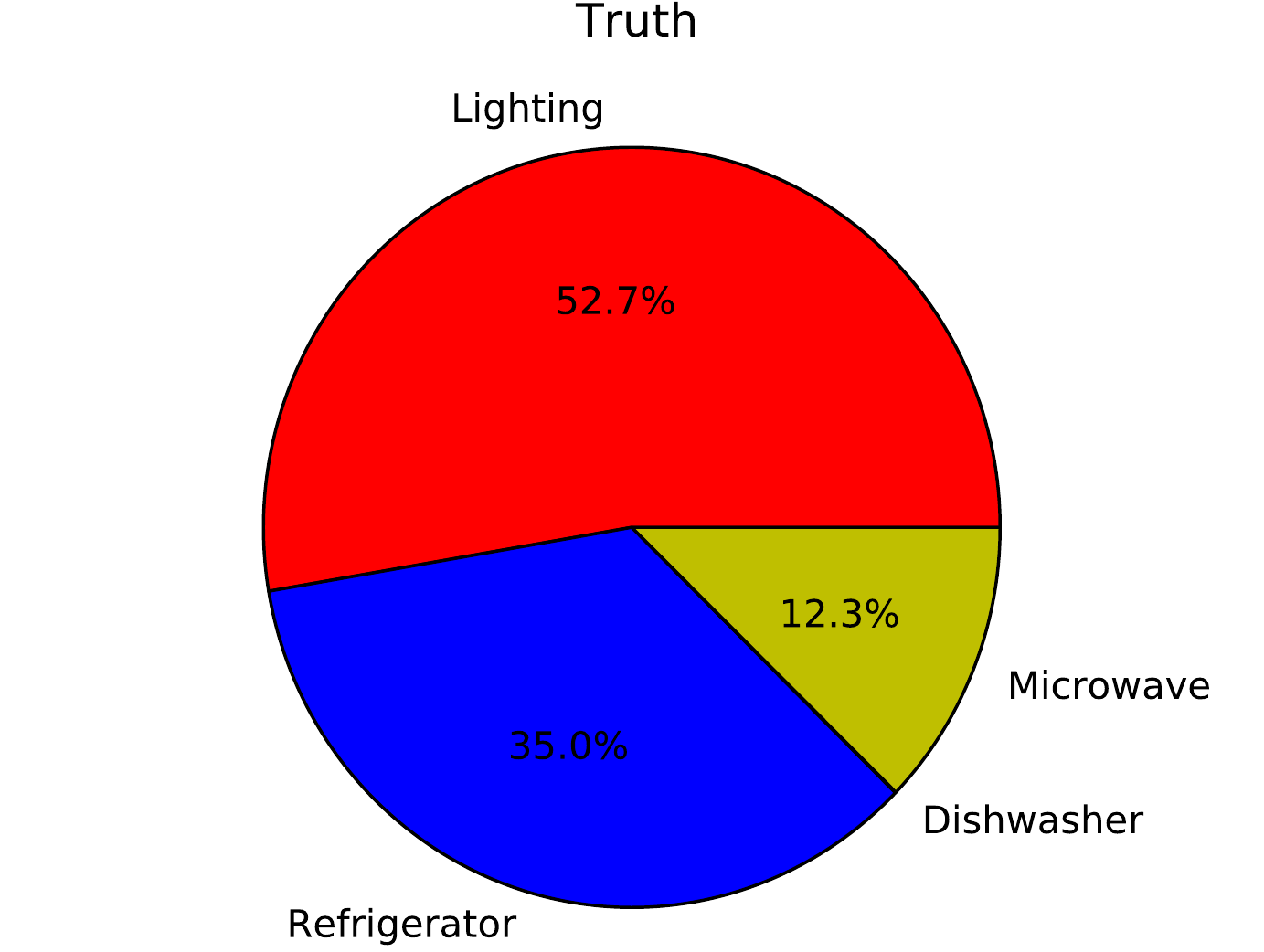}}
    \caption{Estimated total power consumption for a data sequence where both the HDP-HMM and HDP-HSMM approaches performed well.}
    \label{fig:disagg_pies_bothgood}
\end{figure}

Our overall results are summarized in Figure~\ref{fig:disagg_bar} and
Table~\ref{tab:disagg_summary}. Both Bayesian approaches improved upon the
EM-based approach because they allowed flexibility in the device models that
could be fit during inference, while the EM-based approach fixed device model
parameters that may not be consistent across homes. Furthermore, the
incorporation of duration structure and prior information provided a
significant performance increase for the HDP-HSMM approach.  Detailed
performance comparisons between the HDP-HMM and HDP-HSMM approaches can be seen
in Figure~\ref{fig:disagg_seq_breakdown}.  Finally,
Figures~\ref{fig:disagg_pies_different} and~\ref{fig:disagg_pies_bothgood}
shows total power consumption estimates for the two models on two selected data
sequences.

\begin{table}[tp]
    \centering
    \begin{tabular}{|c|c|c|c|}
        \hline
        House & EM FHMM & F-HDP-HMM & F-HDP-HSMM \\\hline
        1 & 46.6\% & 69.0\% & 82.1\% \\
        2 & 50.8\% & 70.7\% & 84.8\% \\
        3 & 33.3\% & 67.3\% & 81.5\% \\
        6 & 55.7\% & 61.8\% & 77.7\% \\
        \hline
        Mean & 47.7\% & 67.2\% & 81.5\% \\
        \hline
    \end{tabular}
    \caption{Performance comparison broken down by house.}
    \label{tab:disagg_summary}
\end{table}

We note that the nonparametric prior was very important for modeling the power
consumption due to lighting. Power modes arise from combinations of lights
switched on in the user's home, and hence the number of levels that are
observed is highly uncertain a priori. For the other devices the number of
power modes (and hence states) is not so uncertain, but duration statistics can
provide a strong clue for disaggregation; for these, the main advantage of our
model is in providing Bayesian inference and duration modeling.

\section{Conclusion}
\label{sec:conclusion}
We have developed the HDP-HSMM and two Gibbs sampling inference algorithms,
the weak limit and direct assignment samplers, uniting
explicit-duration semi-Markov modeling with new Bayesian nonparametric
techniques. These models and algorithms not only allow learning from complex
sequential data with non-Markov duration statistics in supervised and
unsupervised settings, but also can be used as tools in constructing and
performing infernece in larger hierarchical models. We have demonstrated the
utility of the HDP-HSMM and the effectiveness of our inference algorithms with
real and synthetic experiments, and we believe these methods can be built upon
to provide new tools for many sequential learning problems.

\acks{The authors thank J.~Zico Kolter, Emily Fox, and Ruslan Salakhutdinov for
    invaluable discussions and advice. We also thank the anonymous reviewers
    for helpful fixes and suggestions. This work was supported in part by a MURI through
    ARO Grant W911NF-06-1-0076, in part through a MURI through AFOSR Grant
    FA9550-06-1-303, and in part by the National Science Foundation Graduate
    Research Fellowship under Grant No. 1122374.}

% \bibliography{main}
\phantomsection
\bibliography{main}

\appendix

\section{Power Disaggregation Priors}
\label{appendix:power-priors}
We used simple hand-set priors for the power disaggregation experiments in
Section~\ref{sec:power-disaggregation}, where each prior had two free
parameters that were set to encode rough means and variances for each device
mode's durations and emissions. To put priors on multiple modes we used atomic
mixture models in the priors. For example, refrigerators tend to exhibit an
``off'' mode near zero Watts, an ``on'' mode near 100-140 Watts, and a ``high''
mode near 300-400 Watts; we include each of these regimes in the prior by
specifying three sets of hyperparameters, and a state samples observation
parameters by first sampling one of the three sets of hyperparameters uniformly
at random and then sampling observation parameters using those hyperparameters

A comprehensive summary of our prior settings for the Factorial HDP-HSMM are in
Table~\ref{tab:power-disaggregation-priors}. Observation distributions were all
Gaussian with state-specific latent means and fixed variances.  We use
$\GP(\mu_0,\sigma_0^2;\sigma^2)$ to denote a Gaussian observation distribution prior
with a fixed variance of $\sigma^2$ and a prior over its mean parameter that is
Gaussian distributed with mean $\mu_0$ and variance $\sigma_0^2$; i.e., it
denotes that a state's mean parameter $\mu$ is sampled according to $\mu \sim
\mathcal{N}(\mu_0,\sigma_0^2)$ and an observation from that state is sampled
from $\mathcal{N}(\mu,\sigma^2)$. Similarly, we use $\NGP(\alpha,\beta;r)$ to
denote Negative Binomial duration distribution priors where a latent
state-specific ``success'' parameter $p$ is drawn from $p \sim
\text{Beta}(\alpha,\beta)$ and the parameter $r$ is fixed, so that state
durations for that state are then drawn from $\text{NegBin}(p,r)$. (Note
choosing $r=1$ sets a geometric duration class.)

\begin{figure}[tp]
    \centering
    \includegraphics[width=4in]{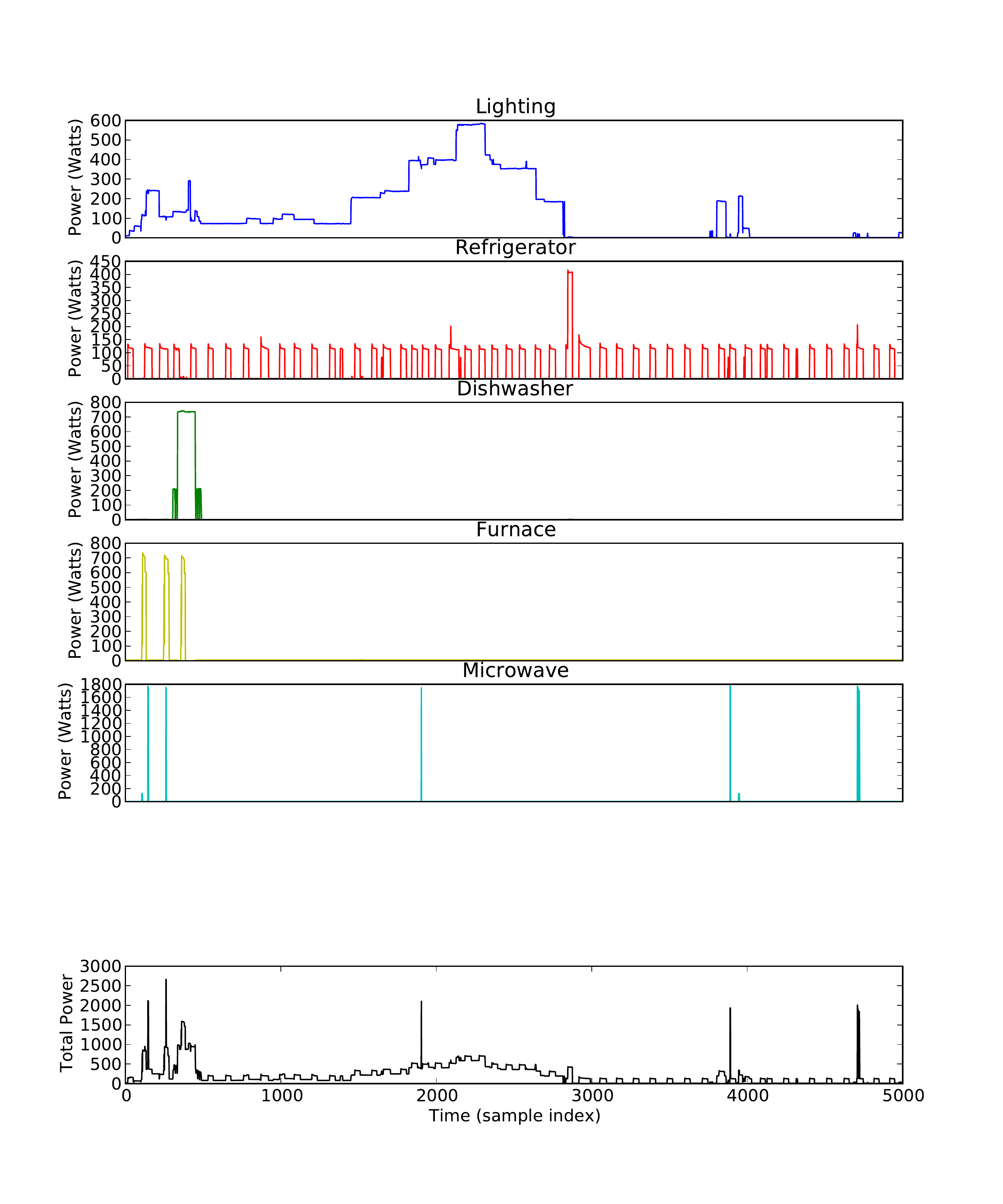}
    \caption{Example real data observation sequences for the power disaggregation experiments.}
    \label{fig:power-disaggregation-real-sequences}
\end{figure}

We set the priors for the Factorial Sticky HDP-HMM by using the same set of
observation prior parameters as for the HDP-HSMM and setting state-specific
sticky bias parameters so as to match the expected durations encoded in the
HDP-HSMM duration priors. For an example of real data observation sequences, see
Figure~\ref{fig:power-disaggregation-real-sequences}.

A natural extension of this model would be a more elaborate hierarchical model
which learns the hyperparameter mixtures automatically from training data. As
our experiment is meant to emphasize the merits of the HDP-HSMM and sampling
inference, we leave this extension to future work.

% \begin{figure}[tp]
%     \centering \includegraphics[width=6in]{figures/disagg_experiments/disagg-priors.pdf}
%     \caption{Samples from simple hand-set appliance priors used in power signal
%         disaggregation experiments. In each plot, each colored line is an
%         independent sample.} \label{fig:power-disaggregation-power-draws}
% \end{figure}

\begin{table}[p]\footnotesize
    \centering
    \begin{tabular}{|c|c|c|c|c|}
        \hline
        \multirow{2}{*}{Device}         & \multicolumn{2}{c}{Base Measures}                                           & \multicolumn{2}{|c|}{Specific States}\\\cline{2-5}
                                        & Observations                          & Durations                           & Observations         & Durations \\\hline\hline

        Lighting                        & $\GP(300,200^2;5^2)$                  & $\NGP(5,220;12)$                    & $\GP(0,1;5^2)$       & $\NGP(5,220;12)$   \\\hline

        \multirow{3}{*}{Refrigerator}   & \multirow{3}{*}{$\GP(110,50^2;10^2)$} & \multirow{3}{*}{$\NGP(100,600;10)$} & $\GP(0,1;5^2)$       & $\NGP(100,600;10)$ \\
                                        &                                       &                                     & $\GP(115,10^2;10^2)$ & $\NGP(100,600;10)$ \\
                                        &                                       &                                     & $\GP(425,30^2;10^2)$ & $\NGP(100,600;10)$ \\\hline

        \multirow{3}{*}{Dishwasher}     & \multirow{3}{*}{$\GP(225,25^2;10^2)$} & \multirow{3}{*}{$\NGP(100,200;10)$} & $\GP(0,1;5^2)$       & $\NGP(1,2000;1)$   \\
                                        &                                       &                                     & $\GP(225,25^2;10^2)$ & $\NGP(100,200;10)$ \\
                                        &                                       &                                     & $\GP(900,200^2;10^2)$& $\NGP(40,500;10)$  \\\hline

        Furnace                         & $\GP(600,100^2;20^2) $                & $\NGP(40,40;10)$                    & $\GP(0,1;5^2)$       & $\NGP(1,50;1)$     \\\hline

        Microwave                       & $\GP(1700,200^2;50^2)$                & $\NGP(200,1;50)$                    & $\GP(0,1;5^2)$       & $\NGP(1,1000;1)$   \\\hline
\end{tabular}
\caption{Power disaggregation prior parameters for each device. Observation priors encode rough power levels that are expected from devices. Duration priors encode duration statistics that are expected from devices.}
\label{tab:power-disaggregation-priors}
\end{table}

\end{document}